\documentclass[aps.prd,twocolumn,floatfix,superscriptaddress,showpacs]{revtex4-1}

\usepackage{graphicx}% Include figure files
\usepackage{bm}% bold math
\usepackage{bbold}
\usepackage{amssymb,amsfonts,amsmath}
\usepackage{hyperref}% add hypertext capabilities
\usepackage{color}
\usepackage{amsfonts}
\usepackage{subfigure}
\usepackage{array}
\usepackage{ulem}

\begin{document}

\preprint{}

\title{Revisiting the Equation of State of Hybrid Stars in the Dyson-Schwinger Equation Approach to QCD}
%
% Force line breaks with \\

\author{Zhan Bai}
\email{baizhan@pku.edu.cn}
\affiliation{Department of Physics and State Key Laboratory of Nuclear Physics and Technology, Peking University, Beijing 100871, China}
\affiliation{Collaborative Innovation Center of Quantum Matter, Beijing 100871, China}
\author{Huan Chen}
%\email{}
\affiliation{School of Mathematics and Physics, China University of Geosciences, Wuhan 430074, China}

\author{Yu-xin Liu}
\email[Corresponding author: ]{yxliu@pku.edu.cn}
\affiliation{Department of Physics and State Key Laboratory of Nuclear Physics and Technology, Peking University, Beijing 100871, China}
\affiliation{Collaborative Innovation Center of Quantum Matter, Beijing 100871, China}
\affiliation{Center for High Energy Physics, Peking University, Beijing 100871, China}

\date{\today}

\begin{abstract}
We investigate the equation of state(EoS) and the effect of the hadron-quark phase transition of strong interaction matter in compact stars.
The hadron matter is described with the relativistic mean field theory,
and the quark matter is described with the Dyson-Schwinger equation approach of QCD.
The complete EoS of the hybrid star matter is constructed with not only the Gibbs construction but also  the 3-window interpolation.
The mass-radius relation of hybrid stars is also investigated.
We find that, although the EoSs of both the hadron matter with hyperon and $\Delta$-baryon
and the quark matter are generally softer than that of the nucleon matter,
the 3-window interpolation construction may provide an EoS stiff enough for a hybrid star with mass exceeding 2$M_{\odot}^{}$ and, in turn, solve the so called ``hyperon puzzle". .
\end{abstract}

%\pacs{25.75.Nq, %Quark deconfinement, quark-gluon plasma production, and phase transitions
%      04.30.Tv, %Gravitational-wave astrophysics
%      26.60.Kp, %Equations of state of neutron-star matter
%      97.60.Bw  %Supernovae
%     }                             % Classification Scheme.

\maketitle

%\tableofcontents

\section{\label{sec:Intro}Introduction}

It has been well known that, when one has an equation of state (EoS) for the dense matter in a star,
one can calculate the mass-radius relation of the star by solving the Tolman-Oppenheimer-Volkov (TOV) equation, and compare the result with astronomical observations.
Compact stars are then regarded as wonderful laboratories in the universe,
which have the extreme condition impossible to reach on earth,
to test the theories for not only cold-dense matter,
but also thermal-dense matter~\cite{Shapiro:1983,Glendenning:2000}, even the existence of critical end point of the QCD phase transitions~\cite{Castillo:2016EPJA}.

The EoS has been known as the essential of astronomical researches (for reviews, see, e.g., Refs.~\cite{Lattimer:2001ApJ,Lattimer:200716PRs,Ozel:2009PRD,Oertel:2017RMP}).
Specifically, the maximum mass of a compact star is highly dependent on the stiffness of the EoS
at high density~\cite{Lattimer:2001ApJ,Lattimer:200716PRs,Ozel:2009PRD,Oertel:2017RMP,Baldo:1997AA,Akmal:1998PRC,Glendenning:2000,Zhou:2004PRC,Li:2008PRC}.
For the ones composed of only proton, neutron and electron (hereafter, we denote it as pure nucleon matter), the EoS can be stiff enough and support a highly massive star.
However, the components at the suprasaturation densities in the core of a compact star is not well determined.
For hadron matter, hyperonic degree of freedom is likely to appear at about
$(2\sim 3)\, \rho_{\textrm{sat}}^{}$ ($\rho_{\textrm{sat}}^{}$ is the saturation nuclear matter density,
$0.153\ \textrm{fm}^{-3}$)~\cite{Bunta:2004PRC,Schulze:2006PRC,Sedrakian:2007PPNP,Shao:2009PRC,Shao:2010PRC,Fortin:2017PRC}.
The appearance of hyperons softens greatly the EoS,
and this contradicts with the observations of the compact stars with mass about $2M_{\odot}^{}$~\cite{Demorest:2010Nature,Antoniadis:2013Science}.
This is called ``hyperon puzzle" in literature (e.g., Refs.~\cite{Lonardoni:2015PRL,Masuda:2016EPJA}).
In addition, the possible appearance of $\Delta$-resonance states (simply referred as $\Delta$-baryons in the follows) might also soften the EoS.
For a long time it is believed that the $\Delta$-baryons will appear in nuclear matter at a density of $\sim 10\rho_{\textrm{sat}}$~\cite{Glendenning:2000},
which is too high even in compact stars.
However, later researches show that with proper symmetry energy and parameterization,
the $\Delta$-baryons may appear at a density relevant in compact stars~\cite{Hu:2003PRC, Schurhoff:2010ApJ,Lavagno:2010PRC, Drago:2014PRC,Cai:2015PRC, Drago:2016EPJA}.
And this formulates the ``$\Delta$ puzzle'', similar to the ``hyperon puzzle.''

One way to solve the ``hyperon puzzle" is the modification of the interactions at high density.
There have been phenomenological models for the matter consisting of nucleons and hyperons and corresponding leptons (hereafter we refer to the star composed of such kind of matter simply as neutron star) which predict neutron stars of mass exceeding $2\, M_{\odot}^{}$
(see, e.g., Refs.~\cite{Lopes:2014PRC,Gomes:2015ApJ,Oertel:2015JPG}).
In microscopic models, there have been relativistic Dirac-Brueckner-Hartree-Fock (DBHF) calculations
with three-body forces~\cite{Katayama:2015PLB},
auxiliary field diffusion Monte Carlo method calculations~\cite{Lonardoni:2013PRC,Lonardoni:2014PRC,Lonardoni:2015PRL}, and so forth.

However, the possible appearance of quark matter should also be taken into consideration,
and is believed as a straight forward way to solve the ``hyperon puzzle" (see, e.g. the discussion in Ref.~\cite{Demorest:2010Nature}).
In the center of a compact star, the baryon density may reach, or even exceed $1\,\textrm{fm}^{-1}$.
At such a high density, the baryons overlap with each other,
and the quark degree of freedom is likely to appear.
The compact star with a quark matter core and a hadron matter mantle is called a ``hybrid star''.
If there is a quark matter core inside the compact star, the EoS of the star matter will also be changed,
and expected to be stiff enough to support a high-mass compact star.

A simple but widely implemented model for the quark matter is the MIT bag model (see, e.g., Refs.~\cite{Chodos:1974PRD,Farhi:1984PRD,Burgio:2002PRC,Nicotra:2006PRD}),
and there have been models beyond the MIT bag model for cold quark matter,
e.g. the Nambu--Jona-Lasino (NJL)  model~\cite{Buballa:1996NPA,Shao:2011PRDa,Shao:2011PRDb,Schertler:1999,Shao:2013PRD,Klahn:2013PRD,Benic:2015AA},
the density-dependent-quark-mass model~\cite{Chakrabarty:1991PRD,Benvenuto:1995PRD,Torres:2013EPL},
the chiral quark meson model~\cite{Zacchi:2015PRD},
quasi-particle model~\cite{Peshier:2000PRC,Tian:2012PRD,Zhao:2015PRD},
and extended confined isospin-density-dependent mass model~\cite{Qauli:2016PRD}, and so on.
However, these models are lack of pronounced quantum chromodynamics (QCD) foundation.
Because of the complexity of the nonlinear and nonperturbative nature of the strong interaction between the quarks,
the EoS for the cold quark matter of compact star in sophisticated QCD approach is then still under investigation.

It has been known that the Dyson-Schwinger equations (DSEs) (see, e.g., Ref.~\cite{Roberts:DSE-BSE}) are almost uniquely a continuum QCD approach which includes both the confinement and the dynamical chiral symmetry breaking (DCSB) features simultaneously~\cite{McLerran:2007NPA},
and are successful in describing  QCD phase transitions and hadron properties (see, e.g., Refs.~\cite{Roberts:DSE-BSE,Qin:2011PRL,Gao:2016PRDa,Gao:2016PRDb,QCDPT-DSE2}).
And the MIT bag model, the NJL model and other phenomenological models can be regarded as the limiting cases of the DSE approach.
We will then, in this paper, implement the DSE approach to describe the quark matter in the way similar to that used in Refs.~\cite{Chen:2011PRD,Chen:2012PRD,Chen:2015PRD,Chen:2016EPJA} in which only the Gibbs construction (see below) is taken to build the EoS of the hybrid star matter.

For the EoS of the matter in hybrid stars, one should take into account the transition from hadron phase to quark phase.
An ideal theory for describing the hadron-quark phase transition is that all the properties are depicted with an unified Lagrangian for the system,
but such a theory has definitely not yet been established.
At present, one has to describe the quark phase and the hadron phase with separate approaches, and derive the complete EoS by construction.

One of the commonly used methods of the construction is the Gibbs construction~\cite{Glendenning:1992PRD,Glendenning:2000}.
It assumes that there is a mixed phase where both quark and hadron phases coexist.
The pressure and chemical potential of each the two phases equate to each other, respectively.
At the same time, though the hadron and quark phases are charged separately,
they neutralize each other and combine to be charge neutral.
The Gibbs construction has been widely taken to calculate the masses of hybrid stars (see, e.g., Refs.~\cite{Burgio:2002PRC,Chen:2011PRD,Chen:2012PRD,Chen:2015PRD,Chen:2016EPJA,Glendenning:1992PRD,Maruyama:2007PRD,Carroll:2009PRC,Weissenborn:2011ApJL,Fischer:2011ApJS,Schulze:2011PRC,Wen:2013PA}).

However, apart from its success, the Gibbs construction has its limitations.
For example, near the hadron-quark phase transition density,
the distance between quarks in one hadron and that in different hadrons are of the same order,
the hadrons should then not be regarded as point particles.
Though the hadron models are accurate for the matter near the saturation density
due to the calibrations coming from plenty of experiments,
they are unreliable and differed from each other greatly in higher density region, the phase transition regions, (see, e.g., Ref.~\cite{Oertel:2017RMP,Xiao:2009PRL}).
Similarly, the quark models are appropriate at extremely high density where perturbative QCD and asymptotic freedom can be applied,
but they are unreliable at lower densities.
Also, in Gibbs construction, both the quark and the hadron matters in the mixed phase region is assumed to be uniform and equilibrium.
However, because of the Coulomb energy, the surface energy and other finite size effects~\cite{Heiselberg:1993PRL, Glendenning:1995PRC, Christiansen:1997PRC, Endo:2006PTP, Maruyama:2007PRD, Yasutake:2014PRC,Wu:2017PRC},
and also due to the non-equilibrium effect,
the EoS should be very different from that coming from the Gibbs construction.
Nevertheless, since we know little about the properties of matter at the high densities,
the discussion about such effects is still limited.

To fix these problems, the 3-window construction model~\cite{Masuda:2013ApJ,Masuda:2013PTEP,Masuda:2016EPJA,Kojo:2015PRD,Kojo:2016EPJA,Baym:2017RPP},  which assumes that with the increasing of density,
there are also 3 regions, has been proposed.
In the 3-window construction, the matter in low density region consists of only separate hadrons which can be approximated as point particles, the quarks (and gluons) are confined inside hadrons and do not play the role to affect the properties of the matter directly.
At high density, the boundaries of hadrons totally disappear and only quark (and gluon) degrees of freedom exist.
In middle density region, the hadron-quark transition region, the boundaries of hadrons gradually disappear, and there is a smooth ``crossover'' but not a sudden change from hadron matter to quark matter.
In the low and high density regions, the hadron and quark models can be applied, respectively.
In the transition region, however,
the EoS is constructed by interpolating the EoSs in the low and the high density regions.

In this paper, we investigate the EoS of compact star matter in a quite large baryon density region and the mass-radius relation of compact stars to solve the ``hyperon puzzle" (as well as the ``$\Delta$ puzzle").
In our investigation, we take the relativistic mean field (RMF) theory for the matter in hadron phase,
and DSE approach for the matter in quark phase.
For the construction of the interplay we implement both the Gibbs construction and the 3-window interpolation.
Our calculations show that, although the EoS of the hadron matter including hyperons and $\Delta$-baryons and that of the quark matter are generally softer than that of pure nucleon matter,
the 3-window interpolation may provide an EoS stiff enough for a hybrid star with mass exceeding $2M_{\odot}^{}$.

This paper is organized as follows.
After this introduction,
we describe briefly the models for the hadron matter and the quark matter in Sec.~\ref{sec:RMF}, Sec.~\ref{sec:quark}, respectively.
In Sec.~\ref{sec:construction}, we describe the Gibbs construction and the 3-window interpolation for the EoS of the hybrid star matter.
In Sec.~\ref{sec:numerical}, we represent the numerical results of the EoS, the mass-radius relation and the component structure of the matter in the hybrid star.
And Sec.~\ref{sec:sum} is for a summary and some remarks.

\section{\label{sec:RMF} Hadron Matter Sector}

In order to calculate the mass-radius relation of the star composed of mainly strong interaction matter,
we need to calculate the EoS of the matter (hadron matter, or quark matter, or their mixture).
For the hadron matter in which the quark degrees of freedom do not appear, we adopt the relativistic mean field theory.

The relativistic mean field (RMF) theory~\cite{Walecka:1974AP,Boguta:1977,Serot:1986ANP} has been known as one of the successful approaches in describing the properties of  compact nuclear matter~\cite{Glendenning:2000,Oertel:2017RMP}.
There are hundreds of parameterization schemes (models) for the RMF model,
which are based on fitting the properties of nuclear matter.
In Ref.~\cite{Dutra:2014PRC}, 263 RMF models are analyzed and shows that, if they satisfy sufficient nuclear constraints, only a small number of them survive.
And a more strict constraint, the stellar constraint, has been added in Ref.~\cite{Dutra:2016PRC}.
Since the inclusion of hyperons will reduce the maximum mass of neutron star,
the EoS of the matter in pure nucleon phase should be stiff enough to support a neutron star over $2\,M_{\odot}^{}$,
otherwise the model will certainly not allow the observed high mass neutron star after the inclusion of hyperons.
In this paper, we will make use of the TW-99 model~\cite{Typel:1999NPA} which satisfies both the nuclear and the stellar constraints~\cite{Dutra:2014PRC,Dutra:2016PRC}.

The Lagrangian of the TW-99 model for the hadron matter including hyperons
and $\Delta$-baryons is written as:
\begin{equation}\label{eqn:Lagrangian}
\begin{split}
\mathcal{L}=\mathcal{L}_{B}^{} + \mathcal{L}_{\textrm{lep}}^{} + \mathcal{L}_{M}^{} + \mathcal{L}_{\textrm{int}}^{} \, ,
\end{split}
\end{equation}
where $\mathcal{L}_{B}^{}$ is the Lagrangian of free baryons.

In this work, we consider not only the baryon octet $p$,$n$,$\Lambda$,$\Sigma^{\pm,0}$ and $\Xi^{-,0}$,
but also the $\Delta$-resonance states $\Delta^{++,+,0,-}$ ($\Delta$-baryons).

The Lagrangian for the baryon octet reads
\begin{equation}
\mathcal{L}_{\textrm{oct}}^{} = \sum_{\textrm{oct}}{\bar{\Psi}_{i}^{}} (i\gamma_{\mu}^{} \partial^{\mu} - m_{i}^{}){\Psi_{i}^{}}\, ,
\end{equation}
while the Lagrangian for the $\Delta$-baryons is
\begin{equation}
\mathcal{L}_{\Delta}^{} = \sum_{\Delta}{\bar{\Psi}_{\Delta \alpha}^{}} (i\gamma_{\mu}^{} \partial^{\mu} - m_{\Delta}^{}){\Psi_{\Delta} ^{\alpha}}\, .
\end{equation}
In the RMF theory, one can neglect all the complexities arising from the spin-3/2 wave function of the $\Delta$-resonances,
and treat them in the same way as for baryon octet except considering the spin degeneracy of a factor 4~\cite{Lavagno:2010PRC,Kolomeitsev:2017NPA}.

$\mathcal{L}_M$ is the Lagrangian of mesons,
\begin{equation}
\begin{split}
\mathcal{L}_{M}^{} = & \frac{1}{2}\left(\partial_{\mu}^{} \sigma \partial^{\mu} \sigma - m_{\sigma}^{2} \sigma^{2} \right) \\
&	-\frac{1}{4} \omega_{\mu\nu}^{} \omega^{\mu\nu} - \frac{1}{2}m_{\omega}^{2} \omega_{\mu}\omega^{\mu} \\
& -\frac{1}{4}\boldsymbol{\rho}_{\mu\nu}^{} \boldsymbol{\rho}^{\mu\nu} -\frac{1}{2}m_{\rho}^{2} \boldsymbol{\rho}_{\mu} \boldsymbol{\rho}^{\mu} \, ,
\end{split}
\end{equation}
where $\sigma$, $\omega_{\mu}^{}$, and $\boldsymbol{\rho}_{\mu}^{}$ are the
isoscalar-scalar, isoscalar-vector and isovector-vector meson field, respectively.
$\omega_{\mu\nu}^{}=\partial_{\mu}\omega_{\nu} - \partial_{\nu}\omega_{\mu}$,
$\boldsymbol{\rho}_{\mu\nu}^{} =\partial_{\mu}\boldsymbol{\rho}_{\nu} - \partial_{\nu}\boldsymbol{\rho}_{\mu}\,$.

The $\mathcal{L}_{\textrm{int}}^{}$ in Eq.~(\ref{eqn:Lagrangian}) is the Lagrangian
 describing the interactions between baryons which are realized by exchanging the mesons:
\begin{equation}
\begin{split}
\mathcal{L}_{\textrm{int}}^{} = & \sum_{B} g_{\sigma B}^{} {\bar{\Psi}_{B}^{}} \sigma {\Psi_{B}^{}}
-g_{\omega B}^{} {\bar{\Psi}_{B}^{}} \gamma_{\mu}^{} \omega^{\mu} {\Psi_{B}^{}} \\
& - g_{\rho B}^{} {\bar{\Psi}_{B}^{}} \gamma_{\mu}^{} \boldsymbol{\tau}_{B}^{} \cdot \boldsymbol{\rho}^{\mu} {\Psi_{B}^{}} \, ,
\end{split}
\end{equation}
where $g_{iB}^{}$ for $i=\sigma$, $\omega$, $\rho$ are the coupling strength parameters between baryons and mesons, which depend on the baryon density.

In some other literatures, the self-interaction of $\sigma$-meson, the cross interaction between different kind mesons and the effect of the isovector-scalar $\delta$-meson are included explicitly (see, e.g., the review in Refs.~\cite{Oertel:2017RMP,Dutra:2014PRC}).
In TW-99 parameterization, however, all these terms are taken zero,
and their effects are represented in the density dependence of the coupling constants.
For nucleons, the coupling constants are
\begin{equation}
{g_{iN}^{}}(\rho_{B}^{}) = {g_{iN}^{}} (\rho_{\textrm{sat}^{}}) f_{i}^{} (x),  \qquad \quad \textrm{for} \quad i=\sigma,\omega, \rho,
\end{equation}
where $\rho_{B}^{}$ is the baryon density, $\rho_{\textrm{sat}}^{}$ is the saturation nuclear matter density  and
$x= {\rho_{B}^{}}/{\rho_{\textrm{sat}}^{}}$.
The density function can be chosen as~\cite{Typel:1999NPA}:
\begin{equation}
\begin{split}
f_{i}^{}(x) & = a_{i}^{} \frac{1+b_{i}^{} (x + d_{i}^{})^{2}}{1 + c_{i}^{}(x+d_{i}^{})^{2}}, \qquad \textrm{for}
\qquad i=\sigma,\omega \, , \\
f_{\rho}^{}(x) & = \textrm{exp}\left[-a_{\rho}^{} (x-1) \right] \, ,
\end{split}
\end{equation}
where the parameters $a_{i}^{}$, $b_{i}^{}$, $c_{i}^{}$, $d_{i}^{}$ and $g_{iN}^{}(\rho_{\textrm{sat}}^{})$
are fixed by fitting the properties of the nuclear matter at the saturation density,
and their values are listed in Table~\ref{tab:gi}.

\begin{table}[htb]
\begin{center}
\caption{Parameters of the mesons and their couplings (taken from Ref.~\cite{Typel:1999NPA}).}\label{tab:gi}
\begin{tabular}{c|ccc}
\hline
Meson i & $\sigma$ & $\omega$ & $\rho$ \\
\hline
~$m_i$(MeV)~ & 550 & 783 & 763\\
$g_{iN}^{}(\rho_{\textrm{sat}}^{})$ & ~$10.72854$~ & ~$13.29015$~ & ~$7.32196$~ \\
$a_{i}^{}$ & ~$1.365469$~ & ~$1.402488$~ & ~$0.515$~ \\
$b_{i}^{}$ & ~$0.226061$~ & ~$0.172577$~ & {} \\
$c_{i}^{}$ & ~$0.409704$~ & ~$0.344293$~ & {} \\
$d_{i}^{}$ & ~$0.901995$~ & ~$0.983955$~ & {} \\
\hline
\end{tabular}
\end{center}
\end{table}

For hyperons, we represent them with the relation between the hyperon coupling and the nucleon coupling as:
$\chi_{\sigma}^{}=\frac{g_{\sigma Y}^{}}{g_{\sigma N}^{}}$,
$\chi_{\omega}^{}=\frac{g_{\omega Y}^{}}{g_{\omega N}^{}}$,
$\chi_{\rho}^{}=\frac{g_{\rho Y}^{}}{g_{\rho N}^{}}$.
On the basis of hypernuclei experimental data, we choose them as those in Refs.~\cite{Glendenning:2000,Dutra:2016PRC}:
$\chi_{\sigma}^{}=0.7$, $\chi_{\omega}^{}=\chi_{\rho}^{}=0.783$.
For the $\Delta$-baryons, we choose the naive coupling constant: $x_{\sigma\Delta}=x_{\omega\Delta}=x_{\rho\Delta}=1$.

The $\mathcal{L}_{\textrm{lep}}^{}$ is the Lagrangian for leptons, which are treated as free fermions:
\begin{equation}
\mathcal{L}_{\textrm{lep}}^{}=\sum_{l}^{} {\bar{\Psi}_{l}^{}}(i\gamma_{\mu}^{} \partial^{\mu} - m_{l}^{}) {\Psi_{l}^{}} \, ,
\end{equation}
and we include only the electron and muon in this paper.

The field equations can be derived by differentiating the Lagrangian.
Under RMF approximation, the system is assumed to be in the static, uniform ground state.
The partial derivatives of the mesons all vanish,
only the 0-component of the vector meson and the 3rd-component of the isovector meson survive and can be replaced with the corresponding expectation values.
The field equations of the mesons are then:
\begin{equation}\label{eqn:sigma}
m_{\sigma}^{2} \sigma =\sum_{B} g_{\sigma B}^{} \langle {\bar{\Psi}_{B}^{}} {\Psi_{B}^{}} \rangle \, ,
\end{equation}
\begin{equation}\label{eqn:omega}
m_{\omega}^{2} \omega_{0}^{} = \sum_{B} g_{\omega B}^{} \langle {\bar{\Psi}_{B}^{}} \gamma_{0}^{} {\Psi_{B}^{}} \rangle \, ,
\end{equation}
\begin{equation}\label{eqn:rho}
m_{\rho}^{2} \rho_{03}^{} = \sum_{B} g_{\rho B}^{} \langle {\bar{\Psi}_{B}^{}} \gamma_{0}^{} \tau_{3B}^{} {\Psi_{B}^{}} \rangle \, ,
\end{equation}
where $\tau_{3B}^{}$ is the 3rd-component of the isospin of baryon $B$.

The equation of motion (EoM) of the baryon is:
\begin{equation}\label{EOM}
\left[\gamma^{\mu} (i\partial_{\mu}^{} - \Sigma_{\mu}^{}) - (m_{B}^{} - g_{\sigma B}^{} \sigma) \right] {\Psi_{B}^{}} = 0 \, ,
\end{equation}
where
\begin{equation}
\Sigma_{\mu}^{} = g_{\omega B}^{} \omega_{\mu} + g_{\rho B}^{} \boldsymbol{\tau}_{B}^{} \cdot \boldsymbol{\rho}_{\mu}^{} + \Sigma_{\mu}^{\textrm{R}}.
\end{equation}
The $\Sigma_{\mu}^{\textrm{R}}$ is called the ``rearrange'' term,
which appears because of the density-dependence of the coupling constant, and reads
\begin{equation}\label{eqn:rearrange}
\begin{split}
\Sigma_{\mu}^{\textrm{R}} = & \frac{j_{\mu}^{}}{\rho} \bigg(\frac{\partial g_{\omega B}^{}} {\partial\rho}\bar{\Psi}_{B}^{} \gamma^{\nu} \Psi_{B}^{} \omega_{\nu}^{} \\
	& + \frac{\partial g_{\rho B}^{}}{\partial\rho}\bar{\Psi}_{B}^{} \gamma^{\nu} \boldsymbol{\tau}_{B}^{} \cdot \boldsymbol{\rho}_{\nu}^{} \Psi_{B}^{}
		-\frac{\partial g_{\sigma B}^{}}{\partial\rho}\bar{\Psi}_{B}^{} \Psi_{B}^{} \sigma \bigg) \, ,
\end{split}
\end{equation}
where $j_{\mu}^{} = \bar{\Psi}_{B}^{} \gamma_{\mu} \Psi_{B}^{}$ is the baryon current.

Under the EoM of Eq.~(\ref{EOM}),
the baryons behave as quasi-particles with effective mass
\begin{equation}\label{eqn:mstar}
m_{B}^{\ast} = m_{B}^{} - g_{\sigma B}^{} \sigma \, ,
\end{equation}
and effective chemical potential:
\begin{equation}\label{eqn:mustar}
\mu_{B}^{\ast} = \mu_{B}^{} - g_{\omega B}^{} \omega_{0}^{} - g_{\rho B}^{} \tau_{3B}^{} \rho_{03}^{} -\Sigma_{\mu}^{\textrm{R}} \, .
\end{equation}

One can then get the baryon (number) density:
\begin{equation}\label{eqn:ni}
\begin{split}
\rho_{B}^{} \equiv \langle {\bar{\Psi}_{B}^{}} \gamma^{0} {\Psi_{B}^{}} \rangle =\gamma_{B}^{}\int\frac{\textrm{d}^3k}{(2\pi)^3} = \gamma_{B}^{}\frac{k_{FB}^{3}}{6\pi^{2}} \, ,
\end{split}
\end{equation}
where $k_{FB}^{} =\sqrt{\mu_{B}^{\ast 2} - m_{B}^{\ast 2}}$ is the Fermi momentum of the particle,
$\gamma_B$ is the spin degeneracy, which is $2$ for the baryon octet and $4$ for the $\Delta$-baryons.
And the scalar density is:
\begin{equation}\label{eqn:nis}
\begin{split}
	\rho_{B}^{s} \equiv & \langle {\bar{\Psi}_{B}^{}} {\Psi_{B}^{}} \rangle =\gamma_{B}^{}\int\frac{\textrm{d}^3k}{(2\pi)^{3}}\frac{m_{B}^{\ast}}{\sqrt{k^{2} + m_{B}^{\ast 2}}}\\
	&=\gamma_{B}^{}\frac{m_{B}^{\ast}}{4\pi^{2}}\bigg[k_{FB}^{} \mu_{i}^{\ast} - m_{B}^{\ast 2} \textrm{ln}\bigg(\frac{k_{FB}^{} + \mu_{B}^{\ast}}{m_{B}^{\ast}}\bigg)\bigg] \, .
\end{split}
\end{equation}

The expression of the density of leptons is the same as those for baryons,
except that the effective mass and the effective chemical potential should be replaced with the corresponding mass and chemical potential of the leptons:
\begin{equation}\label{eqn:nl}
\rho_{l}^{} = \frac{k_{Fl}^{3}}{3\pi^{2}} \, ,
\end{equation}
where $k_{Fl}^{2} = \mu_{l}^{2} - m_{l}^{2}$ for $l=e^{-},\, \mu^{-}$.

The matter in the star composed of hadrons should be in $\beta$-equilibrium.
Since there are two conservative charge numbers: the baryon number and the electric charge number,
all the chemical potential can be expressed with the neutron chemical potential and the electron chemical potential:
\begin{equation}\label{eqn:beta}
\mu_{i}^{} = B \mu_{n}^{} - Q \mu_{e}^{} \, ,
\end{equation}
where $B$ and $Q$ is the baryon number, electric charge number for the particle $i$, respectively.

Then, combining Eqs.~(\ref{eqn:sigma}), (\ref{eqn:omega}), (\ref{eqn:rho}), (\ref{eqn:rearrange}), (\ref{eqn:mstar}), (\ref{eqn:mustar}), (\ref{eqn:ni}), (\ref{eqn:nis}), (\ref{eqn:nl}) and (\ref{eqn:beta}),
together with the charge neutral condition:
\begin{equation}
\rho_{p}^{} + \rho_{\Sigma^{+}}^{} +\rho_{\Delta^{+}}^{} +2\rho_{\Delta^{++}}^{} = \rho_{e}^{} + \rho_{\mu^{-}}^{} + \rho_{\Sigma^{-}}^{}
+\rho_{\Xi^{-}}^{}+\rho_{\Delta^{-}}^{} \, ,
\end{equation}
one can determine the ingredients and the properties of the hadron matter with any given baryon density $\rho_{B}^{}\,$.

The EoS of hadron matter can be calculated from the energy-momentum tensor:
\begin{equation}
T^{\mu\nu} = \sum_{\phi_i} \frac{\partial\mathcal{L}}{\partial(\partial_{\mu}\phi_{i}^{})}\partial^{\nu}\phi_{i}^{} - g^{\mu\nu}\mathcal{L}.
\end{equation}
The energy density $\varepsilon$ is:
\begin{equation}
\varepsilon=\langle T^{00}\rangle =\sum_{i=B,l}\varepsilon_{i}^{} +\frac{1}{2}m_{\sigma}^{2} \sigma^{2} +\frac{1}{2}m_{\omega}^{2} \omega_{0}^{2} + \frac{1}{2}m_{\rho}^{2} \rho_{03}^{2} \, ,
\end{equation}
where the contribution of the baryon $B$ to the energy density is:
\begin{equation}
\begin{split}
\varepsilon_{B}^{} & = \gamma_{B}^{}\int\frac{\textrm{d}^3k}{(2\pi)^3}\sqrt{k^{2} +m_{B}^{\ast 2}}\\
	&= \! \gamma_{B}^{}\frac{1}{4\pi^{2}} \Big[ 2\mu_{B}^{\ast 3} k_{FB}^{} \! - \! m_{B}^{\ast 2} \mu_{B}^{\ast} k_{FB}^{}
\! - \! m_{B}^{\ast 4}\textrm{ln}\Big(\! \frac{\mu_{B}^{\ast} \! + \! k_{FB}^{}}{m_{B}^{\ast}} \! \Big) \Big].
\end{split}
\end{equation}

The contribution of the leptons to the energy density can be written in the similar form as baryons with a spin degeneracy parameter $\gamma_{l}^{} = 2 $,
except that the effective mass and effective chemical potential should be replaced with those of the leptons, respectively.

As for the pressure of the system, we can determine that with the general formula:
\begin{equation}
P =\sum_{i} \mu_{i}^{} \rho_{i}^{} - \varepsilon \, .
\end{equation}

\section{\label{sec:quark} Quark Matter Sector}

To describe the properties of the matter in quark phase, we adopt the DSE approach of QCD~\cite{Roberts:DSE-BSE}.

The starting point of the DSE approach is the gap equation:
\begin{equation}
S(p;\mu)^{-1}=Z_{2}^{} [i\boldsymbol{\gamma} \cdot  \boldsymbol{p} + i\gamma_{4}^{} (p_{4}^{} + i\mu)+m_{q}^{}]+\Sigma(p;\mu),
\end{equation}
where $S(p;\mu)$ is the quark propagator, $\Sigma(p;\mu)$ is the renormalized self-energy of the quark:
\begin{equation}
\begin{split}
\Sigma(p;\mu)=&Z_{1}^{} \int^{\Lambda}\frac{\textrm{d}^{4} q}{(2\pi)^{4}}
          g^{2}(\mu)D_{\rho\sigma}^{} (p-q;\mu)\\
		&\times\frac{\lambda^{a}}{2} \gamma_{\rho} S(q;\mu) \Gamma_{\sigma}^{a}(q,p;\mu),
\end{split}
\end{equation}
where $\int^{\Lambda}$ is the translationally regularized integral,
$\Lambda$ is the regularization mass-scale.
$g(\mu)$ is the strength of the coupling, $D_{\rho\sigma}^{}$ is the dressed gluon propagator,
$\Gamma_{\sigma}^{a}$ is the dressed quark-gluon vertex,
$\lambda^{a}$ is the Gell-Mann matrix, and $m_{q}^{}$ is the current mass of the quark.
In this paper, for simplicity, the current mass of $u$ and $d$ quark is taken to be zero,
and the current mass of $s$ quark is chosen to be $115\;$MeV, by fitting the kaon mass in vacuum~\cite{Chen:2011PRD}.
$Z_{1,2}^{}$ is the renormalization constants.

At finite chemical potential,
the quark propagator can be decomposed according to the Lorentz structure as:
\begin{equation}
\begin{split}
S(p;\mu)^{-1}=&i\boldsymbol{\gamma} \cdot  \boldsymbol{p} A(p^{2}, p \, u, \mu^{2})
+ B(p^{2},p\, u, \mu^{2})\\
		&+i\gamma_{4}^{} (p_{4} + i\mu) C(p^{2},p\, u, \mu^{2}) \, ,
\end{split}
\end{equation}
with $u=(\boldsymbol{0},i\mu)$.

At zero chemical potential, a commonly used ansatz for the dressed gluon propagator and the dressed quark-gluon interaction vertex is:
\begin{equation}
\begin{split}
Z_{1}^{} g^{2} & D_{\rho\sigma}^{}(p-q)\Gamma_{\sigma}^{a}(q,p) \\
		&=\mathcal{G}((p-q)^{2} )D_{\rho\sigma}^{\textrm{free}}(p-q) \frac{\lambda^a}{2}
             \Gamma_{\sigma}^{}(q,p) \, ,
\end{split}
\end{equation}
where
\begin{equation}
D_{\rho\sigma}^{\textrm{free}}(k\equiv p-q)=\frac{1}{k^{2}} \Big( \delta_{\rho\sigma}^{} - \frac{k_{\rho}^{} k_{\sigma}^{}}{k^{2}} \Big) \, ,
\end{equation}
$\mathcal{G}(k^{2})$ is the effective interaction introduced in the model,
and $\Gamma_{\sigma}^{}$ is the quark-gluon vertex.
In this paper, the rainbow approximation of the vertex is adopted,
\begin{equation}
\Gamma_{\sigma}^{}(q,p) = \gamma_{\sigma}^{} \, .
\end{equation}

For the interaction part, we adopt the Gaussian type effective interaction (see, e.g., Refs.~\cite{Qin:2011PRL,Chen:2011PRD,Maris:1997PRC,Alkofer:2002PRD,Chang:2005NPA,Chen:2008PRD}):
\begin{equation}
\frac{\mathcal{G}(k^{2})}{k^{2}}=\frac{4\pi^{2} D}{\omega^{6}}k^{2} \textrm{e}^{-k^{2}/\omega^{2}} \, ,
\end{equation}
where $D$ and $\omega$ are the parameters of the model.
In this paper we take $\omega=0.5\,$GeV and $D=1.0\,\textrm{ GeV}^2$ as the same as in many literatures.

In case of finite chemical potential, an exponential dependence of the ${\mathcal{G}}$ on the chemical potential was introduced in Ref.~\cite{Chen:2011PRD} as:
\begin{equation}\label{eqn:alpha}
\frac{\mathcal{G}(k^{2};\mu)}{k^{2}} = \frac{4\pi^{2} D}{\omega^{6}} \textrm{e}^{-\alpha\mu^{2}/\omega^{2}}k^{2} \textrm{e}^{-k^{2}/\omega^{2}} \, ,
\end{equation}
where $\alpha$ is the parameter controlling the rate for the quark matter to approach the asymptotic freedom.
It is evident that, when $\alpha=0$, it is the same as that at zero chemical potential;
when $\alpha=\infty$, the effective interaction is zero and corresponds to the case of MIT bag model.
We adopt such a model in our calculation in this paper, and for simplicity, we take the same interaction for each flavor of the quarks.

Moreover, Ref.~\cite{Chen:2015PRD} has calculated the properties of quark matter with several different vertex models and shown that the vertex effect can be absorbed into the variation of the parameter $\alpha$.
We then in this paper adopt only the rainbow approximation of the quark-gluon interaction vertex in our calculations.

With the above equations, we can get the quark propagator,
and derive the EoS of the quark matter in the same way as taken in Refs.~\cite{Chen:2008PRD,Qin:2011PRL,Klahn:2010PRC,Gao:2016PRDa}.

The number density of quarks as a function of its chemical potential is:
\begin{equation}
n_{q}^{}(\mu) = 6\int\frac{\textrm{d}^{3}p}{(2\pi)^{3}} f_{q}^{}(|\boldsymbol{p}|;\mu),
\end{equation}
where $f_{q}^{}$ is the distribution function and reads
\begin{equation}
f_{q}^{}(|\boldsymbol{p}|;\mu) = \frac{1}{4\pi}\int_{-\infty}^{\infty}\textrm{d}p_{4}^{} \textrm{tr}_{\textrm{D}}^{}[-\gamma_{4}^{} S_{q}^{}(p;\mu)] \, ,
\end{equation}
where the trace is for the spinor indices.

The pressure of each flavor of quark at zero temperature can be obtained by integrating the number density:
\begin{equation}
P_{q}^{}(\mu_{q}^{}) = P_{q}^{}(\mu_{q,0}^{}) + \int_{\mu_{q,0}^{}}^{\mu_{q}^{}} \textrm{d}\mu n_{q}^{}(\mu) \, .
\end{equation}

The total pressure of the quark matter is the sum of the pressure of each flavor of quark:
\begin{equation}
P_{Q}^{}(\mu_{u}^{},\mu_{d}^{},\mu_{s}^{}) = \sum_{q=u,d,s}\tilde{P}_{q}^{}(\mu_{q}^{}) - B_{\textrm{DS}}^{} \, ,
\end{equation}
\begin{equation}
\tilde{P}_{q}^{}(\mu_{q}^{})\equiv \int_{\mu_{q,0}^{}}^{\mu_{q}^{}}\textrm{d}\mu n_{q}^{}(\mu) \, ,
\end{equation}
\begin{equation}\label{eqn:B_DS}
B_{\textrm{DS}}^{} \equiv -\sum_{q=u,d,s} P_{q}^{}(\mu_{q,0}^{}) \, .
\end{equation}

Theoretically, the starting point of the integral $\mu_{q,0}^{}$ can be any value,
in this paper we take $\mu_{q,0}^{}=0$.
For the value of $B_{\textrm{DS}}^{}$, a discussion can be seen in Ref.~\cite{Chen:2016EPJA}.
Here we adopt the ``steepest-descent" approximation
and take $B_{\textrm{DS}}^{}=90\textrm{ MeV fm}^{-3}$~\cite{Haymaker:1991RNCSIF,Chen:2008PRD,Chen:2011PRD}.

The quark matter in a compact star should also be in $\beta$-equilibrium and electric charge neutral,
so we have:
\begin{equation}
\mu_{d}^{} = \mu_{u}^{} + \mu_{e}^{} = \mu_{s}^{} \, ,
\end{equation}
\begin{equation}
\frac{2\rho_{u}^{} - \rho_{d}^{} - \rho_{s}^{}}{3} - \rho_{e}^{} - \rho_{\mu^{-}}^{} = 0 \, .
\end{equation}
And we have the baryon density and chemical potential as:
\begin{equation}
\rho_{B}^{} = \frac{1}{3}(\rho_{u}^{} + \rho_{d}^{} + \rho_{s}^{} ) \, ,
\end{equation}
\begin{equation}\label{muB}
\mu_{B}^{} = \mu_{u}^{} + 2\mu_{d}^{} \, .
\end{equation}

Therefore, we can calculate the properties of the quark matter with a given baryon chemical potential (baryon density).

\section{\label{sec:construction} Construction of the Complete Equation of State}

\subsection{Gibbs Construction}

After having the EoSs of both the hadron matter and the quark matter,
we derive the complete EoS of the hybrid star matter by construction.

A widely used construction is the Gibbs construction~\cite{Glendenning:2000,Glendenning:1992PRD}.
It assumes that there is a mixed phase in a density region
in which both quarks and hadrons coexist.
Because of the conservations of the baryon number and the electric charge number,
the baryon chemical potential and the charge chemical potential are the same, respectively, in both the quark and the hadron phases.
In hadron matter, the baryon chemical potential $\mu_{B}^{}$ is the same as the chemical potential of neutron $\mu_{n}^{}$. In quark matter, $\mu_{B}^{}$ is defined by Eq.~(\ref{muB}).
Since electron carries zero baryon number and one minus electric charge number,
we have $\mu_{Q}^{}= - \mu_{e}^{}$,
where $\mu_{Q}^{}$ is the charge chemical potential and $\mu_{e}^{}$ is the electron chemical potential.

In the mixed region, the pressure of the two phases are the same.
And though the two phases may not be charge neutral separately,
there still exists a global electric charge neutral constraint.
If we define the quark fraction $\chi$ with $\chi \in [0,\, 1]$,
the phase transition condition can be expressed as:
\begin{equation}\label{eqn:condition1}
p_{H}^{}(\mu_{n}^{}, \mu_{e}^{}) = p_{Q}^{} (\mu_{n}^{}, \mu_{e}^{} ) \, ,
\end{equation}
\begin{equation}\label{eqn:condition2}
(1-\chi)\rho_{H}^{c} (\mu_{n}^{}, \mu_{e}^{}) + \chi \rho_{Q}^{c}(\mu_{n}^{} , \mu_{e}^{}) = 0 \, ,
\end{equation}
where $p_{H}^{}$ and $p_{Q}^{}$ is the pressure of the matter in hadron, quark phase, respectively,
which is a function of both $\mu_{n}^{}$ and $\mu_{e}^{}$.
And $\rho_{H}^{c}$ and $\rho_{Q}^{c}$ is the electric charge density of the two phases, respectively,
whose sum equates to zero.

Combining Eqs.~(\ref{eqn:condition1}) and (\ref{eqn:condition2}),
together with the field equations of the two phases in Sec.~\ref{sec:RMF} and Sec.~\ref{sec:quark},
we can get the $\mu_{n}^{}$ and $\mu_{e}^{}$ with a given quark fraction $\chi$.
Then, we can calculate the pressure, the energy density and the baryon density of the two separate phases at the phase equilibrium state.
The energy density and the baryon density of the mixed phase (in fact at the phase equilibrium state) are the corresponding superpositions of the contributions from the two phases, and read
\begin{equation}
\varepsilon_{M}^{}= \chi \varepsilon_{Q}^{} (\mu_{n}^{},\mu_{e}^{})
 + (1-\chi)\varepsilon_{H}^{} (\mu_{n}^{},\mu_{e}^{}) \, ,
\end{equation}
\begin{equation}
\rho_{M}^{} = \chi \rho_{Q}^{} (\mu_{n}^{}, \mu_{e}^{}) + (1 - \chi) \rho_{H}^{} (\mu_{n}^{} , \mu_{e}^{}) \, ,
\end{equation}
where the subscripts $M,\,Q$ and $H$ correspond to the mixed, quark and hadron phase, respectively.
The pressure of the mixed phase is just the pressure of each of the two phases:
\begin{equation}
p_{M}^{} = p_{H}^{} = p_{Q}^{} \, .
\end{equation}

\begin{figure}[htb]
\includegraphics[width=0.45\textwidth]{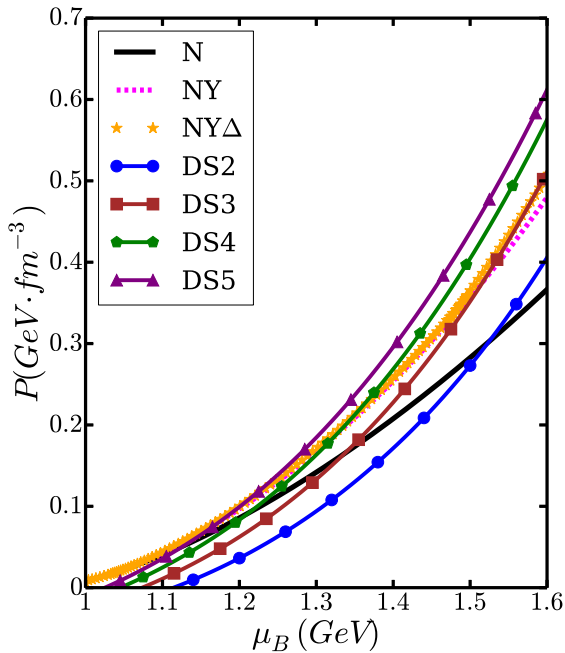}
\caption{Calculated result of the relation between the pressure and the baryon chemical potential of the
matter with different particle components.
The solid, dashed curve corresponds to that of the hadron matter without, with hyperons, respectively.
The star curve denotes that of the hadron matter with both hyperons and $\Delta$-baryons.
The line marked with DS$\alpha$ stands for the result of the pure quark phase with parameter $\alpha$ in Eq.~(\ref{eqn:alpha}).}
\label{fig:p-muB}
\end{figure}

The calculated relation between the pressure and the baryon chemical potential of different phases is
shown in Fig.~\ref{fig:p-muB}.
It has been well known that, under the scheme of Gibbs construction,
the phase transition occurs only if there is a cross point between the $P$--$\mu_{B}^{}$ curves of the quark matter and the hadron matter,
since both the pressure and the chemical potential of the two phases equate to each other in the phase coexistence region (more concretely, at the equilibrium state of the phase transition).
From Fig.~\ref{fig:p-muB} we can notice that, there exists always a cross point between the $P$--$\mu_{B}^{}$ curve of the quark matter with different parameters and the curve of the hadron matter without hyperons, but only some of the quark matter curves (with larger $\alpha$, i.e., weaker couplings) cross the curve of the hadron matter with hyperons and that with both hyperons and $\Delta$-baryons.
It indicates that through the hadron-quark phase transition,  the hadron matter including hyperons can only change to the quark matter with weak couplings.
Therefore  to include the hadron-quark phase transition effects in the hybrid star matter,
only the nucleon matter
was taken into account in Ref.~\cite{Chen:2011PRD} (with the hadron model used in Ref.~\cite{Chen:2011PRD},
when hyperons are included, the phase transition from hadron to quark cannot happen even for large $\alpha$).
Our result manifests that, to get massive compact star, it is naturally not necessary to take the hybrid matter whose hadron matter sector includes hyperons and $\Delta$-baryons under the Gibbs construction  into account in the follows.
Recalling the scheme and the numerical result, one can know that the Gibbs construction takes only the hadron-quark phase transition at the critical state(s) (or phase equilibrium state(s)) into account but does not consider the effects of the phase transitions occurring at different states (for example, different densities) explicitly, which does not match the phase coexistence feature of the first order phase transition at high density (see, e.g., Refs.~\cite{Qin:2011PRL,Gao:2016PRDa,Gao:2016PRDb,QCDPT-DSE2,Masuda:2016EPJA,Masuda:2013ApJ,Masuda:2013PTEP,Kojo:2015PRD,Kojo:2016EPJA}).

\subsection{3-window Interpolation}

In the scheme of the 3-window interpolation construction, as the baryon density increases,
the compact star matter goes through 3 regions.
At low density, the matter is in hadron phase composed of hadrons which are approximated as point particles.
At high density, quarks, the components of hadrons, are no longer confined, so that the properties of the matter are governed by the quark degrees of freedom.
In the middle density,
there should be a transition from hadron matter to quark matter,
where hadrons percolate, and the boundary of any hadron gradually disappears.

The EoSs of both the hadron and the quark phases are based on models.
The hadron model results are accurate near the saturation density,
but differ greatly in high density region.
While the quark models are appropriate in extremely high density region and lose accuracy at low density.
In the transition region,
neither the hadron model nor the quark model represent the nature individually.
Therefore, an interpolation between the quark and the hadron phases should be taken.

Here we adopt the $\varepsilon$-interpolation as a function of the baryon number density as that in Ref.~\cite{Masuda:2013PTEP}, which reads explicitly as:
\begin{equation}\label{eqn:interpolation}
\varepsilon(\rho) = f_{-}^{}(\rho) \varepsilon_{H}^{}(\rho) + f_{+}^{}(\rho) \varepsilon_{Q}^{}(\rho) \, ,
\end{equation}
\begin{equation} \label{eqn:fHfQ}
f_{\pm}^{} = \frac{1}{2}\bigg(1\pm\textrm{tanh}\bigg(\frac{\rho - \bar{\rho}}{\Gamma}\bigg)\bigg) \, ,
\end{equation}
where $\varepsilon_{H}^{}$ and $\varepsilon_{Q}^{}$ are the energy density of the hadron matter and that of the quark matter, respectively.
$\bar{\rho}$ and $\Gamma$ are parameters describing the ``center" density and the width of the transition region, respectively.
In the transition region, the hadron and quark matter may not distribute uniformly,
the $\varepsilon$ is then an approximation of the total energy density.

\begin{figure}[htb]
\includegraphics[width=0.45\textwidth]{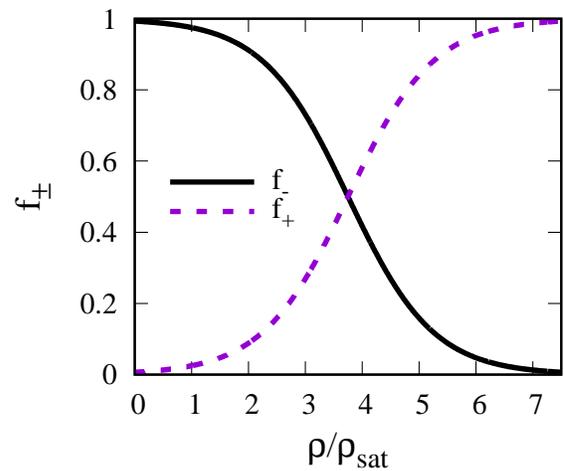}
\caption{Variation behaviors of the interpolation functions $f_{\pm}^{}$ with respect to the baryon number density (in unit of the saturation nuclear matter density $\rho_{\textrm{sat}}^{} $).  }
\label{fig:fHfQ}
\end{figure}

Note that the interpolating function in Eq.~(\ref{eqn:fHfQ})
is different from the $\chi$ in Gibbs construction. The later is the volume fraction of quark matter and characterizes the dependence of the EoS of the matter at the equilibrium state of the phase transition on that of the related two phases (hadron or quark) in a specified region which is determined by solving the coupled equations.
The former, the interpolating function $f_{\pm}^{}$  in the 3-window model,
characterizes the dependence on the hadron (or quark) EoS at any possible density but not only those in a specified region.
The variation behavior of the $f_{\pm}^{}$ in terms of the baryon number density (with parameters $\bar{\rho} = 3.75\,\rho_{\textrm{sat}}^{}$ and $\Gamma = 1.5\, \rho_{\textrm{sat}}^{}$) is shown in Fig.~\ref{fig:fHfQ}.
It can be easily seen that at very low density,
the dependence on the hadron EoS approaches 1 and that on the quark EoS approaches 0,
while at extremely high density the EoS approaches the quark EoS rather than the hadron's.
Moreover, it has been well known that the transition from hadron to quark in high density region is a first order phase transition, whose phase coexistence region exhibits obvious non-uniform, anisotropic and non-equilibrium effects. The 3-window interpolation construction can be interpreted as an approximation for these first order phase transition effects.

The pressure of the transition region can be determined with the thermodynamic relation:
\begin{equation}  \label{eqn:P-Interpolation}
P = \rho^{2} \frac{\partial(\varepsilon/\rho)}{\partial\rho} \, ,
\end{equation}
and the baryon chemical potential $\mu = (\varepsilon + P)/\rho$.

\section{\label{sec:numerical}Numerical Results and Discussions}

\subsection{Equation of State}

\begin{figure}[htb]
\includegraphics[width=0.46\textwidth]{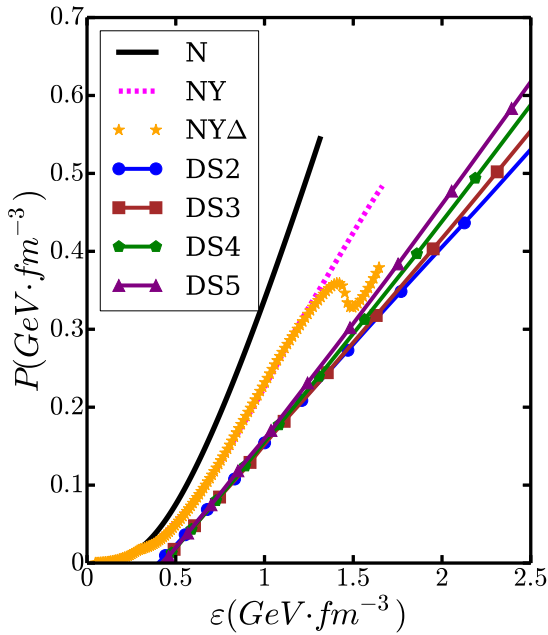}
\caption{Calculated EoSs of the hadron matter with different particle compositions and the pure quark matter via different parameters in the interaction kernel.
The solid curve and dashed curve correspond to that of the hadron phase without and with hyperons, respectively.
The star curve denotes that of the hadron phase with both hyperons and $\Delta$-baryons.
The DS$\alpha$ stands for the result of the pure quark phase with the parameter in Eq.~(\ref{eqn:alpha}) taking a value $\alpha$. }
\label{fig:p-e}
\end{figure}

The calculated results of the relation between the pressure and the energy density (the EoS in convention) of the pure hadron matter and that of the pure quark matter are shown in Fig.~\ref{fig:p-e}.
It is apparent that the hadron matter without hyperons and $\Delta$-baryons has the stiffest EoS,
while all the quark EoSs with different values for the parameter $\alpha$ are softer than all the hadron EoSs, with or without hyperons and $\Delta$-baryons.
The inclusion of hyperons softens the EoS of hadron matter,
and including the $\Delta$-baryons does not change the EoS significantly at low density.
However, at high density, the EoS of the matter with $\Delta$-baryons becomes non-monotonic,
which is in accordance with some of the previous results (see, e.g., Refs.~\cite{Lavagno:2010PRC,Zhu:2016PRC}).
Since a non-monotonic EoS means that the matter in the star is unstable,
and there's no physical solution for the neutron star under the non-monotonic region of the EoS,
we will then take into account only the EoS in the density region in which the EoS of the hadron matter with $\Delta$-baryons is monotonic.

\begin{figure}[htb]
\includegraphics[width=0.45\textwidth]{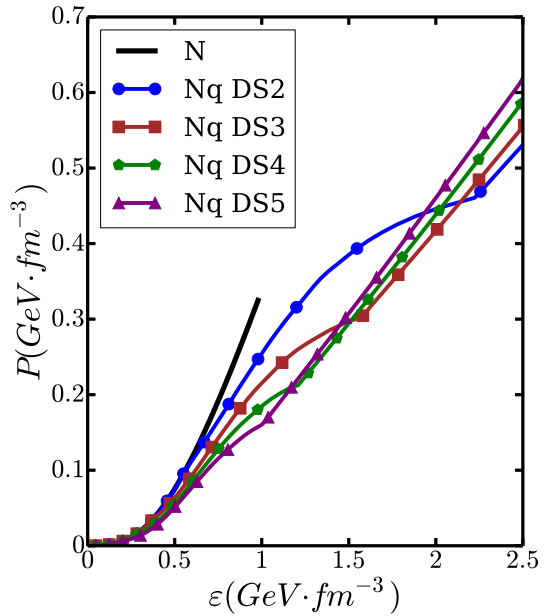}
\caption{Calculated EoSs of the hybrid star matter under the Gibbs construction and that of the pure hadron matter without including hyperons and $\Delta$-baryons (in solid line).
The lines with different symbols display the EoSs of the hybrid matter built via the Gibbs construction.
The Nq DS$\alpha$ marks the one that the quark sector is described with parameter $\alpha$ in Eq.~(\ref{eqn:alpha}) in the DSE approach of QCD.}
\label{fig:eosGNq}
\end{figure}

\begin{figure}[htb]
\includegraphics[width=0.48\textwidth]{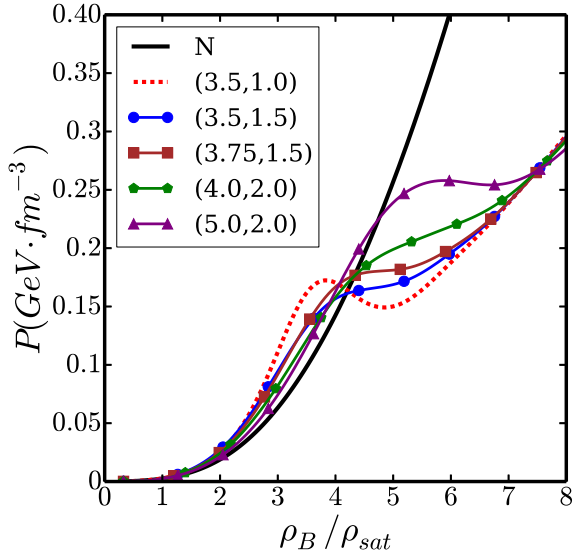}
\caption{Calculated $P$--$\rho_{B}^{}$ curves of the pure nucleon matter (solid line) and the corresponding hybrid star matter in the 3-window interpolation construction with several sets of parameters $\bar{\rho}$ and $\Gamma$ (lines with different symbols).
The quantities of the quark matter sector in the interpolation are fixed via the DSE approach with parameter  $\alpha=2$, where $\alpha$ is defined in Eq.~(\ref{eqn:alpha}). }
\label{fig:rhobar-Gamma}
\end{figure}

The calculated results of the EoSs of the hybrid star matter under the Gibbs construction with quark phase fixed via different parameters in the DSE approach of QCD are shown in Fig.~\ref{fig:eosGNq}.
We take Nq DS$\alpha$ to denote the result of the hybrid matter including the quark phase described with the DSE approach and parameter $\alpha$ in Eq.~(\ref{eqn:alpha}).

Recalling the scheme of the Gibbs construction, one can know that the point at which the curve is not smooth corresponds to the appearance of quark matter(the lower unsmooth point) and the disappearance of hadron matter(the upper unsmooth point),
where the quark fraction $\chi$ equates 0, 1, respectively.
The Fig.~\ref{fig:eosGNq} shows apparently that the phase transition happens
at a lower density in case of larger value of $\alpha$.
This results from the fact that with the increasing of $\alpha$, the interaction (or correlation) between quarks becomes weaker, it is then easier for the quarks to get deconfined from a hadron.
One can also notice that, although the EoS of the pure quark phase with large $\alpha$ is stiffer at high  density,
the EoS of the hybrid star matter in the phase transition region is softer,
because the EoS transits from a stiff nucleon EoS to a stiff quark EoS rather rapidly.

\begin{figure}[htb]
\includegraphics[width=0.43\textwidth]{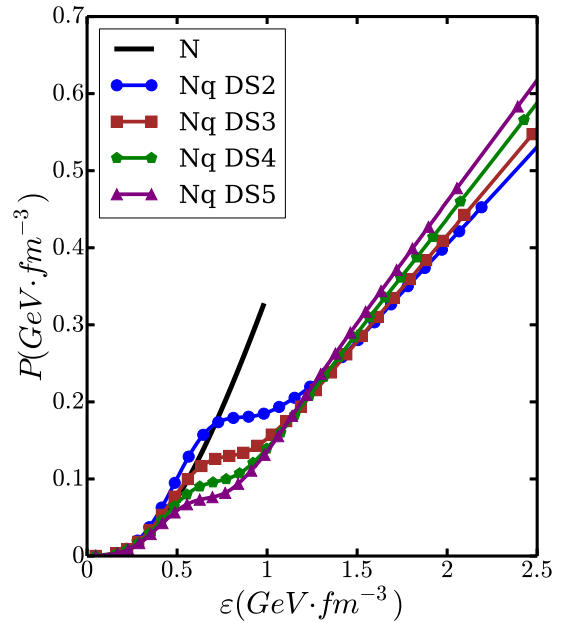}
\caption{Calculated EoSs of the hybrid star matter under the 3-window interpolation construction with   parameters $(\bar{\rho}, \, \Gamma) = (3.75, \, 1.5)\, \rho_{\textrm{sat}}^{}$ (lines with symbols) and that of the hadron matter without including hyperons and $\Delta$-baryons (solid line).
The Nq DS$\alpha$ marks the one that the quark sector is fixed with parameter $\alpha$ in Eq.~(\ref{eqn:alpha}) in the DSE approach. }
\label{fig:eos3Nq}
\end{figure}
\begin{figure}[!htb]
\includegraphics[width=0.43\textwidth]{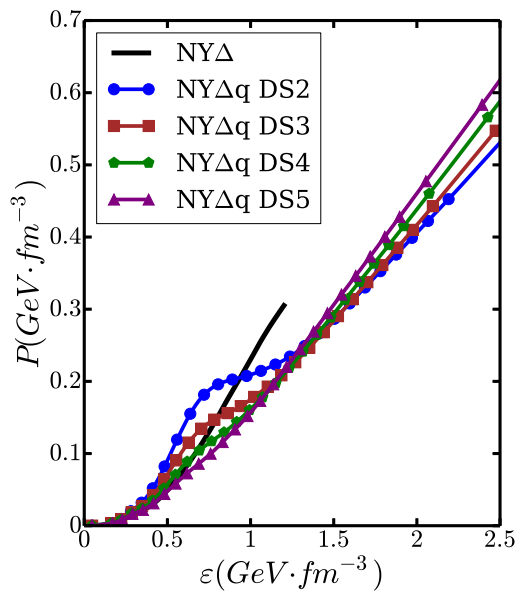}
\caption{The same as Fig.~\ref{fig:eos3Nq} except for that
the hadron sector of the hybrid matter is the one including hyperons and $\Delta$-baryons.} \label{fig:eos3NYDq}
\end{figure}

From Eqs.~(\ref{eqn:interpolation}) and (\ref{eqn:fHfQ}) one can know that, to implement the 3-window interpolation scheme to construct the EoS of the hybrid star matter from those of the hadron matter and the quark matter, one needs the interpolation function in terms of the ``center" density $\bar{\rho}$ and the width $\Gamma$ of the transition region.
We should then fix the parameters $\bar{\rho}$ and $\Gamma$ at first in practical calculations.
The calculated results of the relation between the pressure and the baryon number density ($P$--$\rho_{B}^{}$ curves) under several sets of  $(\bar{\rho},\Gamma)$ values and the corresponding curve of the pure nucleon matter is illustrated in Fig.~\ref{fig:rhobar-Gamma}.

It can be easily seen from Fig.~\ref{fig:rhobar-Gamma} that in case of narrow width $\Gamma$ of the transition region (e.g., the line with filled circles for $(\bar{\rho},\Gamma) = (3.5,\, 1.0)\, \rho_{\textrm{sat}}^{}$)
and quite large central density $\bar{\rho}$ (e.g., the line with up-triangles for $(\bar{\rho},\Gamma) = (5.0,\, 2.0)\, \rho_{\textrm{sat}}^{}$), the constructed EoSs with the 3-window interpolation scheme are non-monotonic, which do not match the general behavior of the EoS of a stable state.
Another constraint on the choice of the $(\bar{\rho},\Gamma)$ is the consistence of the interpolated EoS
with the EoS of the hadron matter at saturation density which is believed to be accurate and requires  $f_{-}^{}(\rho_{\textrm{sat}}^{}) \ge 0.99$ (with $f_{-}^{}(\rho)$ defined in Eq.~(\ref{eqn:fHfQ})).
Considering these two aspects, we take $(\bar{\rho},\Gamma)=(3.75,1.5)\, \rho_{\textrm{sat}}^{}$
in the rest of our calculations in this paper.

The calculated results of the EoSs of the hybrid star matter under the 3-window interpolation construction with parameters $(\bar{\rho}, \, \Gamma) = (3.75, \, 1.5)\, \rho_{\textrm{sat}}^{}$ and the hadron sector without or with the inclusion of the hyperons and $\Delta$-baryons are illustrated in Fig.~\ref{fig:eos3Nq},  Fig.~\ref{fig:eos3NYDq}, respectively.

Comparing with the results in Gibbs construction shown in Fig.~\ref{fig:eosGNq},
one can notice that there is no clear starting and ending points of the phase transition in the interpolation scheme.
It indicates that the 3-window interpolation construction provides a more smooth transition of the EoSs in different phases, i.e., represents the phase coexistence nature of the first order phase transition at high density really explicitly.

More significant difference between the results under the 3-window interpolation and the Gibbs construction is the stiffness of the EoS.
Even though the EoS of the quark phase is generally softer than that of the hadron phase,
the EoS of the hybrid star matter under the 3-window interpolation construction can be stiffer than the hadron EoS in the transition region,
especially for the hadron matter with the inclusion of hyperons and $\Delta$-baryons.
This provides a promise to get large mass hybrid star whose composing matter includes hyperons and $\Delta$-baryons.

\begin{figure}[htb]
\includegraphics[width=0.45\textwidth]{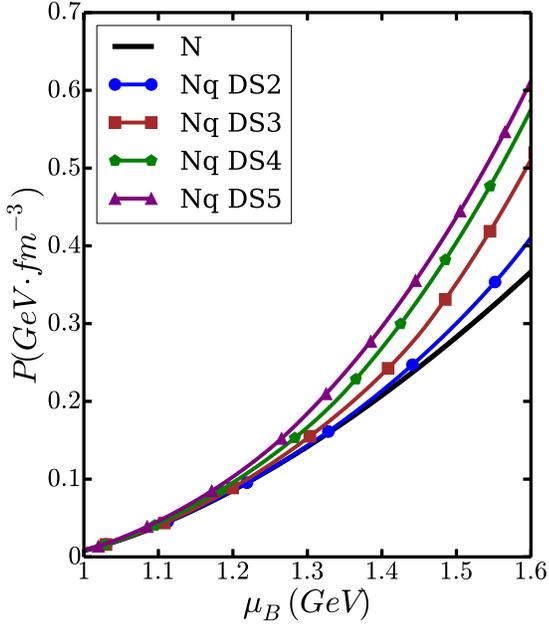}
\caption{Calculated $P$--$\mu_{B}^{}$ curves of the hybrid star matter under the Gibbs construction
(lines with different symbols) and that of the hadron phase without including hyperons and $\Delta$-baryons (solid line).
The Nq DS$\alpha$ marks the one in which the property of the quark sector is described with the parameter $\alpha$ in Eq.~(\ref{eqn:alpha}) in the DSE approach.}
\label{fig:pmuBGNq}
\end{figure}
\begin{figure}[htb]
\includegraphics[width=0.45\textwidth]{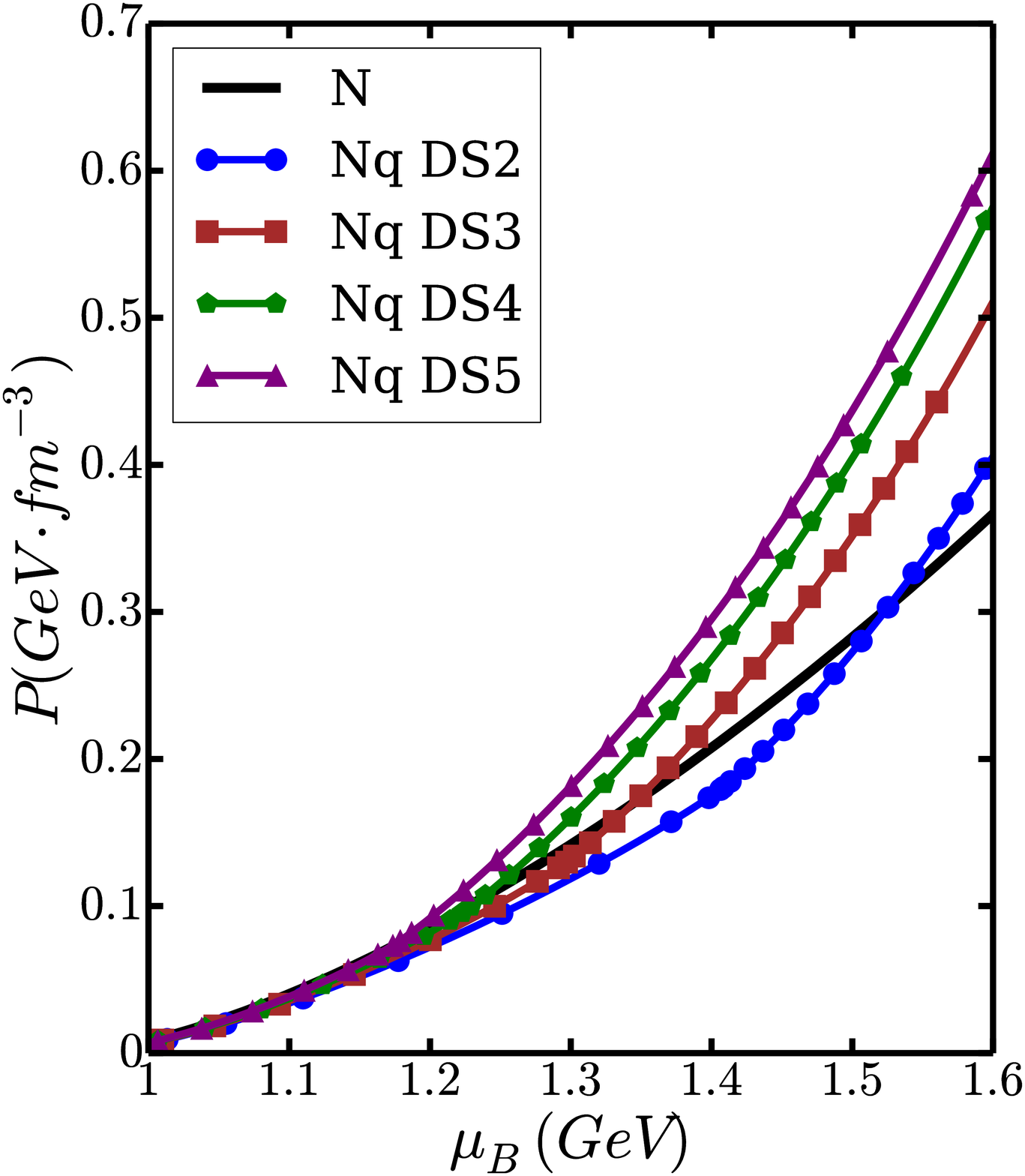}
\caption{Calculated $P$--$\mu_{B}^{}$ curves of the hybrid star matter under the 3-window interpolation construction with 	parameters $(\bar{\rho}, \, \Gamma) = (3.75, \, 1.5)\, \rho_{\textrm{sat}}^{}$ (lines with symbols) and that of the hadron matter without including hyperons and $\Delta$-baryons (solid line).
The Nq DS$\alpha$ marks the one that the quark sector is described with parameter $\alpha$ in Eq.~(\ref{eqn:alpha}) in the DSE approach. }\label{fig:pmuB3Nq}
\end{figure}
\begin{figure}[!htb]
\includegraphics[width=0.45\textwidth]{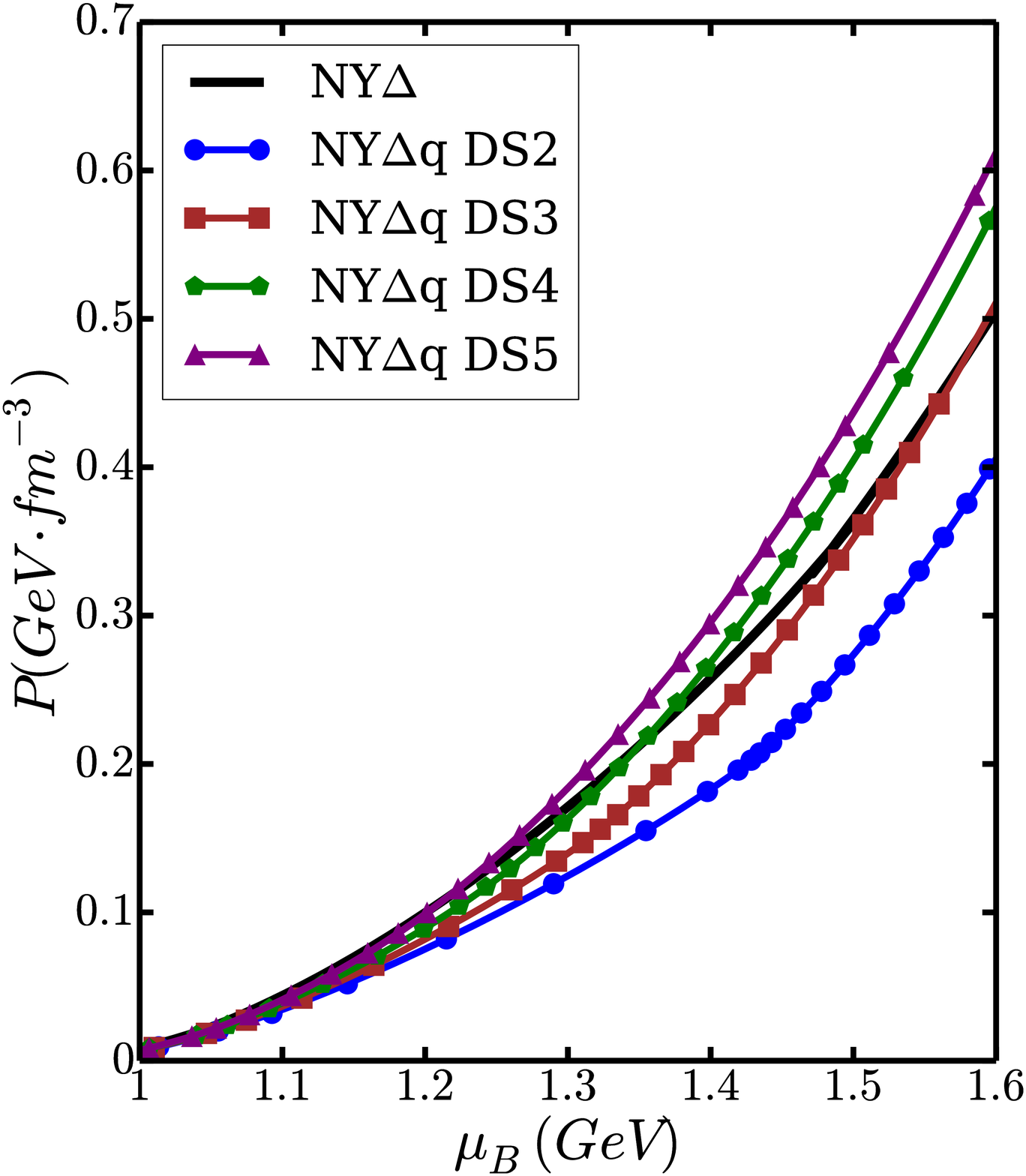}
\caption{The same as Fig.~\ref{fig:pmuB3Nq} except for that
the hadron sector of the hybrid matter is the one including hyperons and $\Delta$-baryons.}\label{fig:pmuB3NYDq}
\end{figure}

\begin{figure}[htb]
\includegraphics[width=0.45\textwidth]{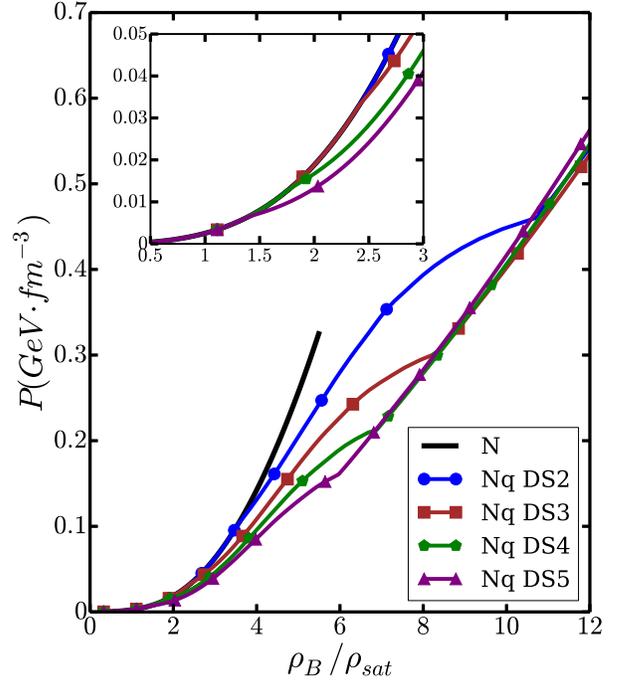}
\caption{Calculated relation between the pressure and the baryon number density ($P$--$\rho_{B}^{}$ curve) of the hybrid matter under the Gibbs construction (lines with symbols)
and that of the hadron matter without including hyperons and $\Delta$-baryons (solid line).
The Nq DS$\alpha$ marks the one that the quark sector is described with parameter $\alpha$ in Eq.~(\ref{eqn:alpha}) in the DSE approach.}
\label{fig:pnBGNq}
\end{figure}

\begin{figure}[htb]
\includegraphics[width=0.45\textwidth]{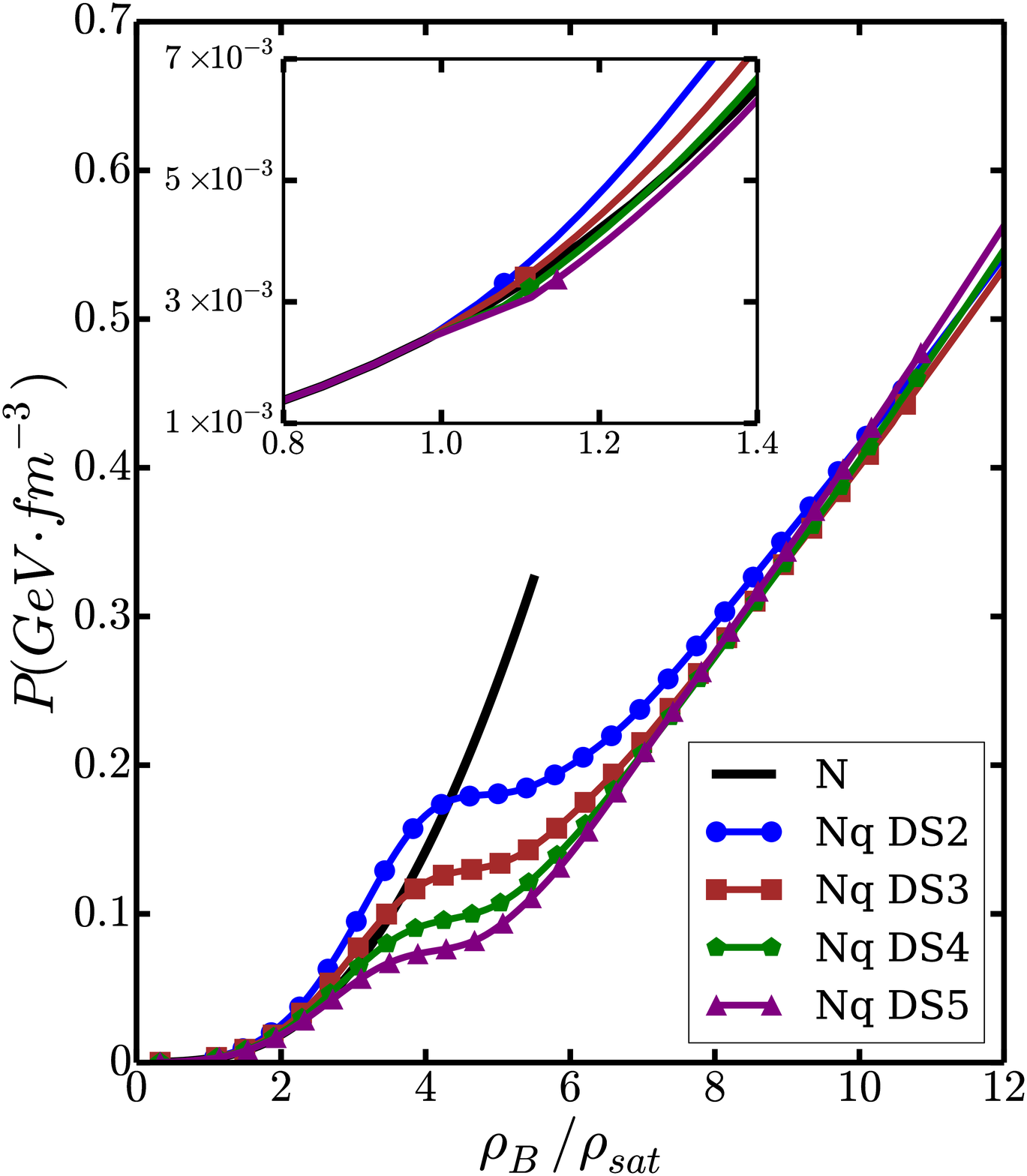}
\caption{Calculated $P$--$\rho_{B}^{}$ curves of the hybrid star matter under the 3-window interpolation construction with 	parameters $(\bar{\rho}, \, \Gamma) = (3.75, \, 1.5)\, \rho_{\textrm{sat}}^{}$ (lines with symbols) and that of the hadron matter without including hyperons and $\Delta$-baryons (solid line).
The Nq DS$\alpha$ marks the one that the quark sector is described with parameter $\alpha$ in Eq.~(\ref{eqn:alpha}) in the DSE approach. }
\label{fig:pnB3Nq}
\end{figure}

\begin{figure}[htb]
\includegraphics[width=0.45\textwidth]{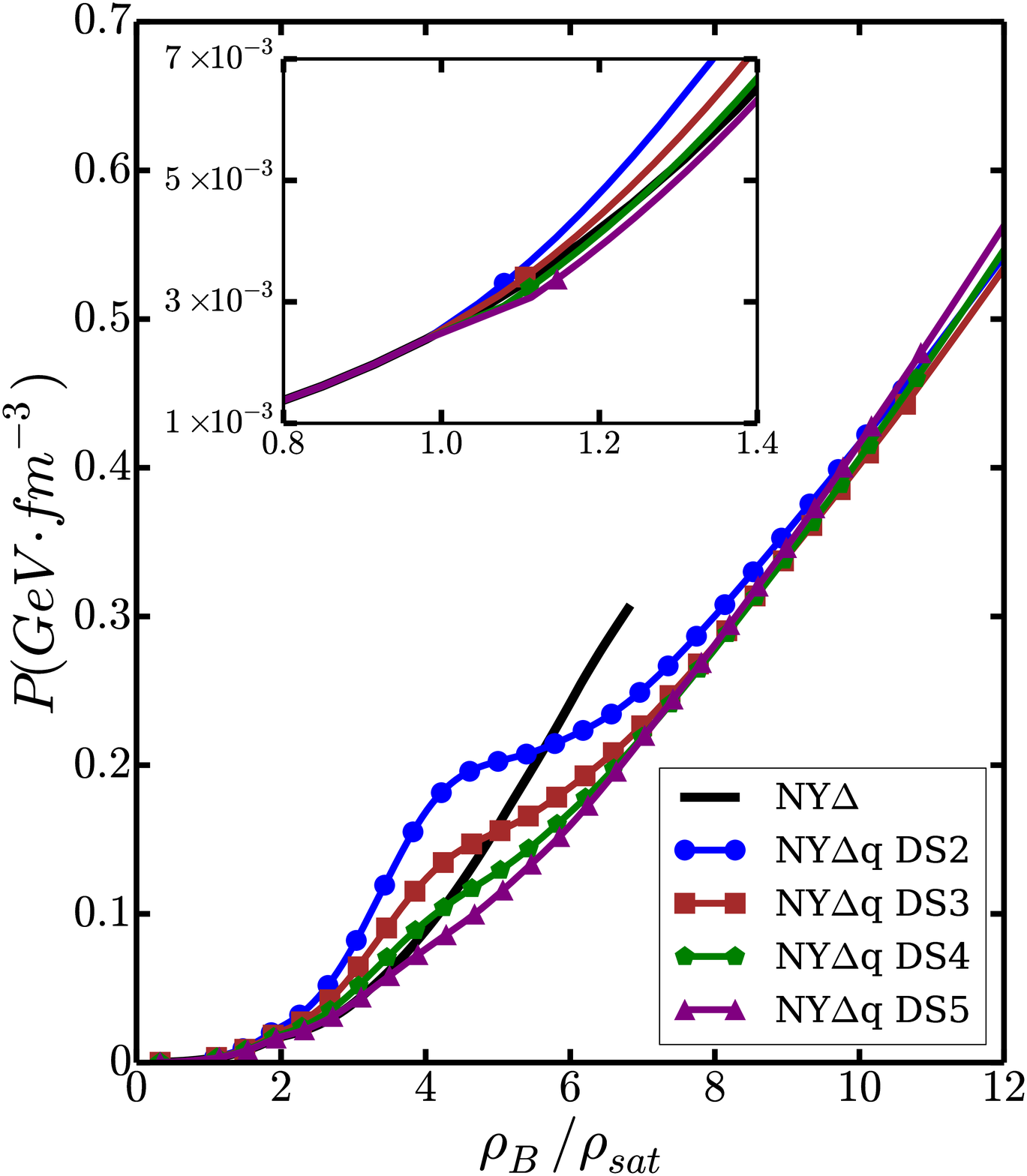}
\caption{The same as Fig.~\ref{fig:pnB3Nq} except for that
the hadron sector of the hybrid matter is the one including hyperons and $\Delta$-baryons.}
\label{fig:pnB3NYDq}
\end{figure}

The calculated results of the relation between the pressure and the baryon chemical potential ($P$--$\mu_{B}^{}$ curves) under different construction schemes are shown in Fig.~\ref{fig:pmuBGNq}, Fig.~\ref{fig:pmuB3Nq} and Fig.~\ref{fig:pmuB3NYDq}.
It is known that, for a uniform and equilibrium phase transition,
the pressure of the mixed phase should be generally greater than not only that of the hadron matter
but also that of the quark matter at a given chemical potential,
and therefore the mixed phase is energy favourable.
The results obtained from the Gibbs construction displayed in Fig.~\ref{fig:pmuBGNq} represents such a feature clearly.

Recalling the scheme of the 3-window interpolation (Eqs.~(\ref{eqn:interpolation}), (\ref{eqn:fHfQ}) and (\ref{eqn:P-Interpolation})), one can know that the pressure of the hybrid star matter reads
\begin{eqnarray}
P & = & - \varepsilon + \rho \frac{\partial \varepsilon}{\partial \rho}  \nonumber \\
& = & f_{-}^{} \Big{[}  - \varepsilon_{H}^{} + \rho \frac{\partial \varepsilon_{H}^{}}{\partial \rho} \Big{]}
  + f_{+}^{} \Big{[}  - \varepsilon_{Q}^{} + \rho \frac{\partial \varepsilon_{Q}^{}}{\partial \rho} \Big{]}
\nonumber \\
& & + \rho\varepsilon_{H}^{} \frac{\partial f_{-}^{}}{\partial \rho}
+ \rho\varepsilon_{Q}^{} \frac{\partial f_{+}^{}}{\partial \rho}  \, . \nonumber
\end{eqnarray}
It manifests that the total pressure is not only a superposition of the pressures of the matter in the two phases with fraction factor $f_{\mp}^{}$ but also other quite complicated terms.
The pressure of the hybrid matter in the phase transition region is then not necessarily larger than both the hadron and quark matters', since the EoSs of not only the hadron but also the quark matters are ``unreliable'' individually.
In other word, the condition mentioned in the last paragraph is not required. However, the pressure of the pure hadron matter at small $\mu_{B}^{}$ and that of the pure quark matter at large $\mu_{B}^{}$ should still be the larger, respectively. Our numerical results shown in Fig.~\ref{fig:pmuB3Nq} and Fig.~\ref{fig:pmuB3NYDq} demonstrate such a characteristic very well.

To show the characteristics of the EoS of the hybrid star matter more explicitly, we display the calculated results of relation between the pressure and the baryon density ($P$--$\rho_{B}^{}$ curves) under different construction schemes in Fig.~\ref{fig:pnBGNq}, Fig.~\ref{fig:pnB3Nq} and Fig.~\ref{fig:pnB3NYDq}.
For the result given in the Gibbs construction shown in Fig.~\ref{fig:pnBGNq},
it is clear that with the increasing of parameter $\alpha$,
phase transition starts and ends at a lower density,
the same as we have seen in Fig.~\ref{fig:eosGNq}.
For $\alpha=2$ case, the phase transition starts at around $4\rho_{\textrm{sat}}^{}$ and ends at around $10\rho_{\textrm{sat}}^{}$.
Such a ending density is too high in the compact star matter.
It manifest that, in $\alpha=2$ case, there will not be a pure quark core inside the compact star
(will be discussed further later).

The Figs.~\ref{fig:pnB3Nq} and \ref{fig:pnB3NYDq} represent obviously that, for the 3-window interpolation,
every constructed $P$--$\rho_{B}^{}$ relation of the hybrid matter starts to deviate
from that of the pure hadron matter at around saturation density.
This is a direct demonstration of the principle of the interpolation scheme.
However, different from the Gibbs construction,
this deviation does not correspond to the sudden appearance of quark matter with considerable fraction and the occurrence of the phase transition.
It means only that from this density, the hadrons cannot be regarded as point particles
and the effect of the quarks inside hadrons begins to play the role.

\subsection{Mass-Radius Relation}

The mass-radius relation of compact stars can be calculated
by solving the Tolman-Oppenheimer-Volkov (TOV) equation:
\begin{equation}
\frac{\textrm{d}P}{\textrm{d} R} = -\frac{G}{ R^2} \big{(} m(R) + 4\pi P R^{3} \big{)} (\varepsilon + P)
\Big{(} 1 - 2 G \frac{m(R)}{R} \Big{)}^{-1} \, ,
\end{equation}
where $G$ is the gravitational constant and $m(R)$ is the mass inside radius $R$:
\begin{equation}
m(R) = \int_{0}^{R} 4\pi r^{2} \varepsilon \textrm{d}r \, .
\end{equation}
Then taking the EoS as input,
one can integrate the TOV equation from inside out to get the mass and radius of the star with a given center density.

\begin{figure}[htb]
\includegraphics[width=0.45\textwidth]{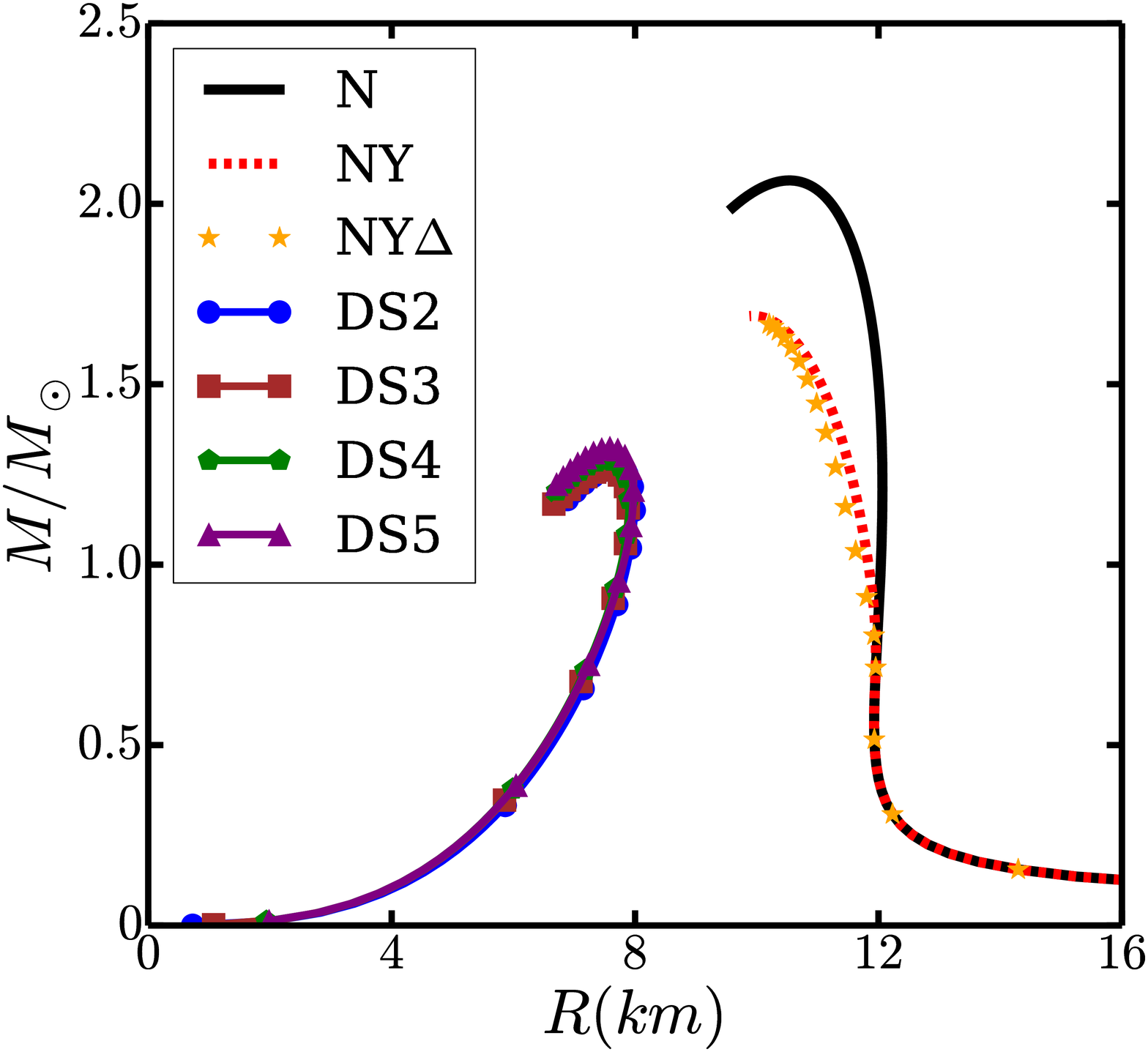}
\caption{Calculated mass-radius relation of a compact star with pure hadron matter or pure quark matter.
The DS$\alpha$ marks the one that the quark matter is described with parameter $\alpha$ in Eq.~(\ref{eqn:alpha}) in the DSE approach of QCD. }\label{fig:mr}
\end{figure}

The obtained mass-radius relation for pure hadron star and pure quark star is shown in Fig.~\ref{fig:mr}.
It is evident that the neutron star consists of purely nucleon matter has the largest maximum mass,
while the inclusion of hyperons and $\Delta$-baryons greatly reduces the maximum mass.
Since the EoS of the hadron matter including both hyperons and $\Delta$-baryons
is not so different from the EoS of the matter including hyperons but without $\Delta$-baryons,
the maximum mass is nearly the same in the two cases.
This confirms the so called ``hyperon puzzle" and the ``$\Delta$ puzzle".
Meanwhile the maximum mass of pure quark star is much lower,
and the radius is also much smaller for quark stars.
Moreover setting different values to the quark parameter $\alpha$ does not change the mass-radius of quark star significantly.

\begin{figure}[htb]
\includegraphics[width=0.45\textwidth]{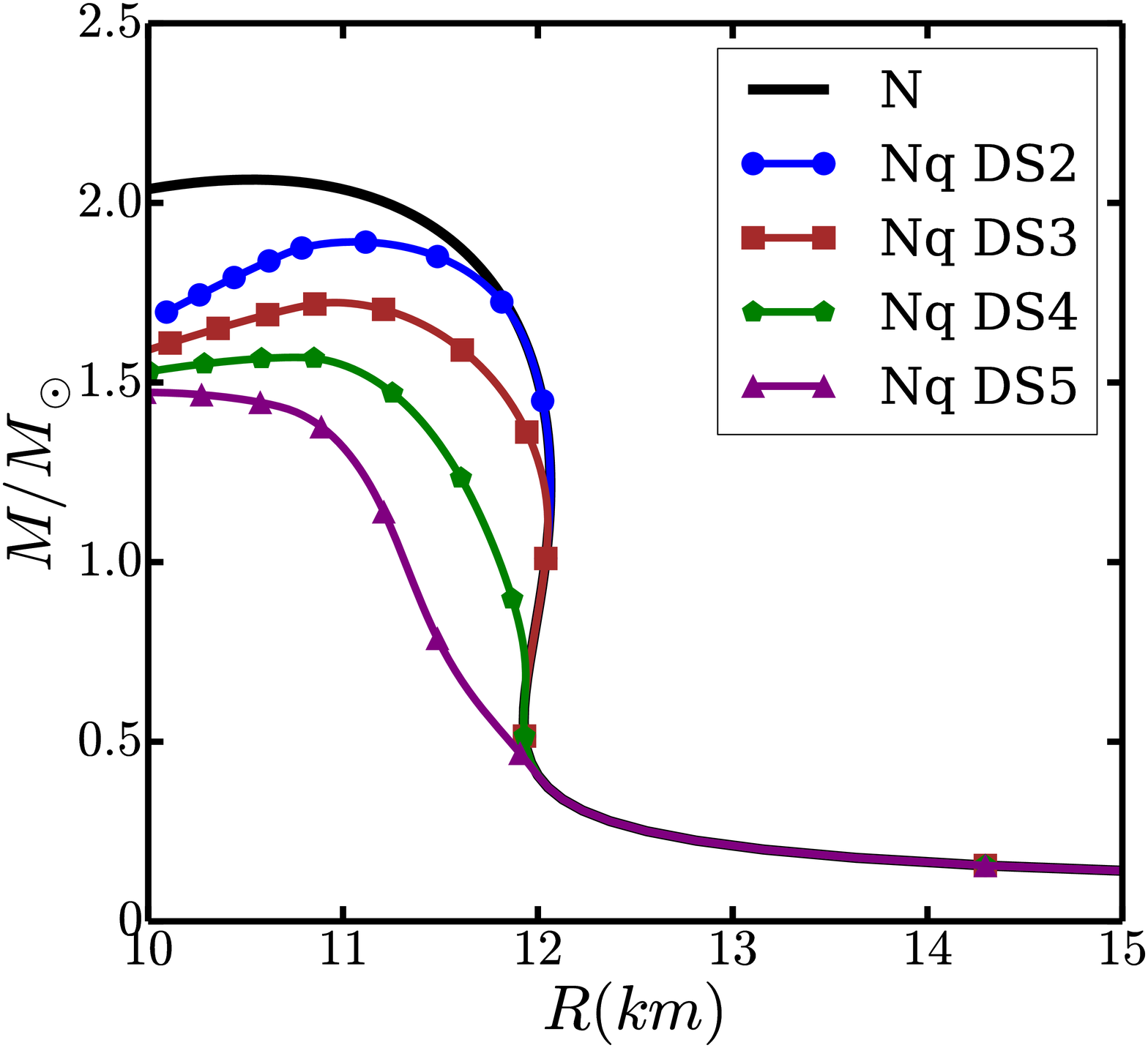}
\caption{Calculated mass-radius relation of a neutron star without the inclusion of hyperons (solid line) and those of hybrid stars with the EoS of the composing matter being fixed using the Gibbs construction (lines with symbols).
The Nq DS$\alpha$ notation marks the one that the quark sector is described with parameter $\alpha$ in Eq.~(\ref{eqn:alpha}) in the DSE approach of QCD. }\label{fig:mrGNq}
\end{figure}

The obtained mass-radius relation of the pure nucleon star whose composing matter does not include either hyperons or $\Delta$-baryons and the corresponding hybrid star under the Gibbs construction is shown in Fig.~\ref{fig:mrGNq}.
The solid line in the figure is the result of the pure nucleon star without the hadron-quark phase transition.
It is apparent that the maximum mass of the pure nucleon star in this case is $2.06\, M_{\odot}^{}$, and the corresponding radius is $10.5\, \textrm{km}$.

Fig.~\ref{fig:mrGNq} manifests clearly that including the quark degrees of freedom via the Gibbs construction reduces the maximum mass of the hybrid star.
For the stars with lower mass, the curves of the mass-radius relation of the hybrid stars are the same as the pure nucleon star's,
because for such stars, the center density is below the hadron-quark phase transition threshold.
In more details, the results of the hybrid star with EoS of the quark matter in DSE approach (DS$\alpha$) with a lager $\alpha$ deviate from that of the nucleon star with small radius drastically.
The maximum mass of the hybrid star drops from $1.89\, M_{\odot}^{}$ in DS2 to $1.47\, M_{\odot}^{}$ in DS5,
and the corresponding radius decreases from $11.05\, \textrm{km}$ in DS2 to $9.96 \, \textrm{km}$ in DS5.
Notice that, in case of  DS2,
the hybrid star reaches its maximum mass soon after the phase transition begins,
and then drops while the mass of the nucleon star is still increasing.
Therefore, the corresponding radius is larger in the DS2 case.

\begin{figure}[htb]
\includegraphics[width=0.45\textwidth]{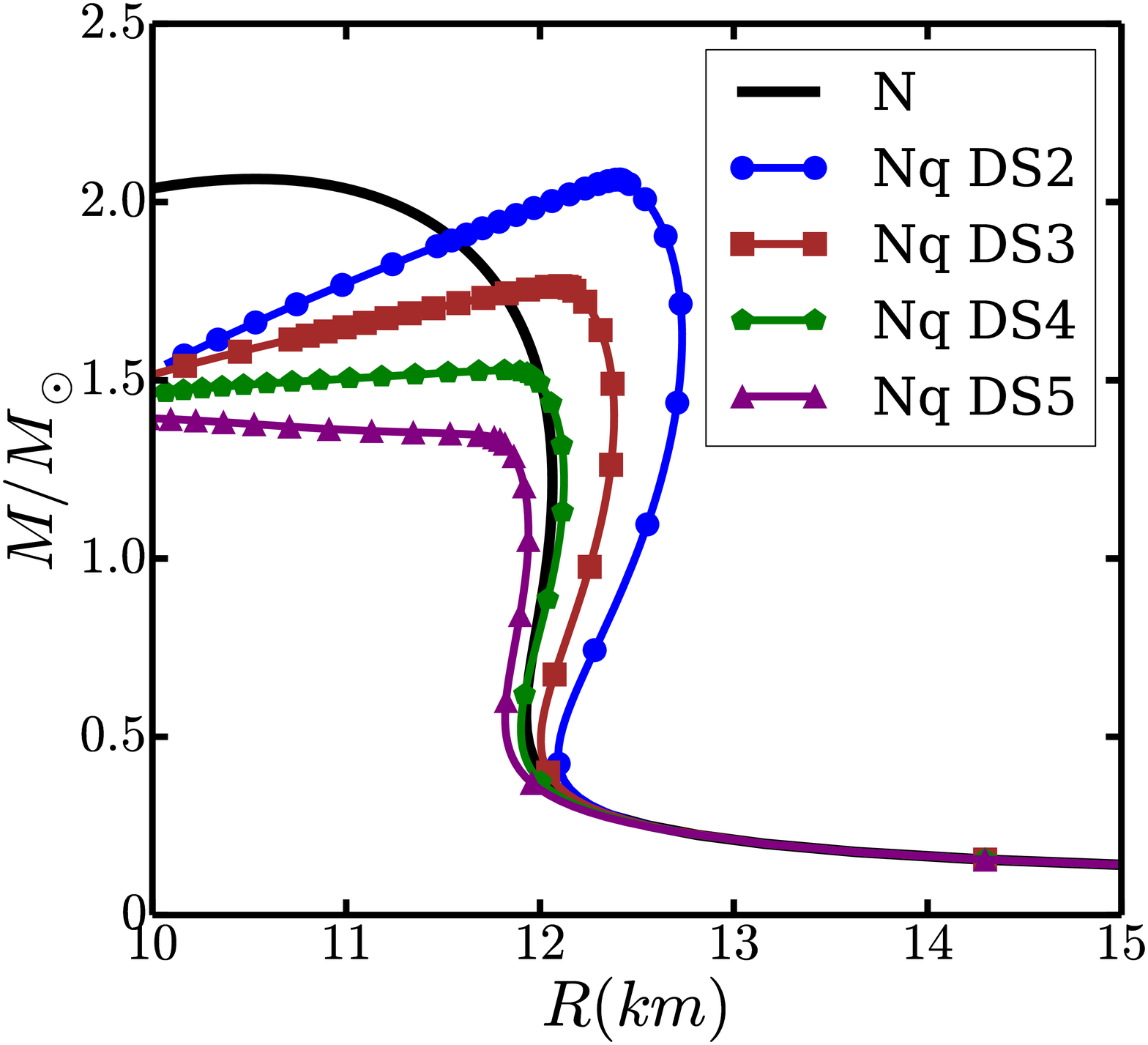}
\caption{Calculated mass-radius relation of a neutron star without the inclusion of hyperons (solid line) and those of hybrid stars with the EoS of the composing matter being constructed using the 3-window interpolation with   parameters $(\bar{\rho}, \, \Gamma) = (3.75, \, 1.5)\, \rho_{\textrm{sat}}^{}$ (lines with symbols).
The Nq DS$\alpha$ notation marks the one that the quark sector is described with parameter $\alpha$ in Eq.~(\ref{eqn:alpha}) in the DSE approach of QCD.
}\label{fig:mr3Nq}
\end{figure}

\begin{figure}[htb]
\includegraphics[width=0.45\textwidth]{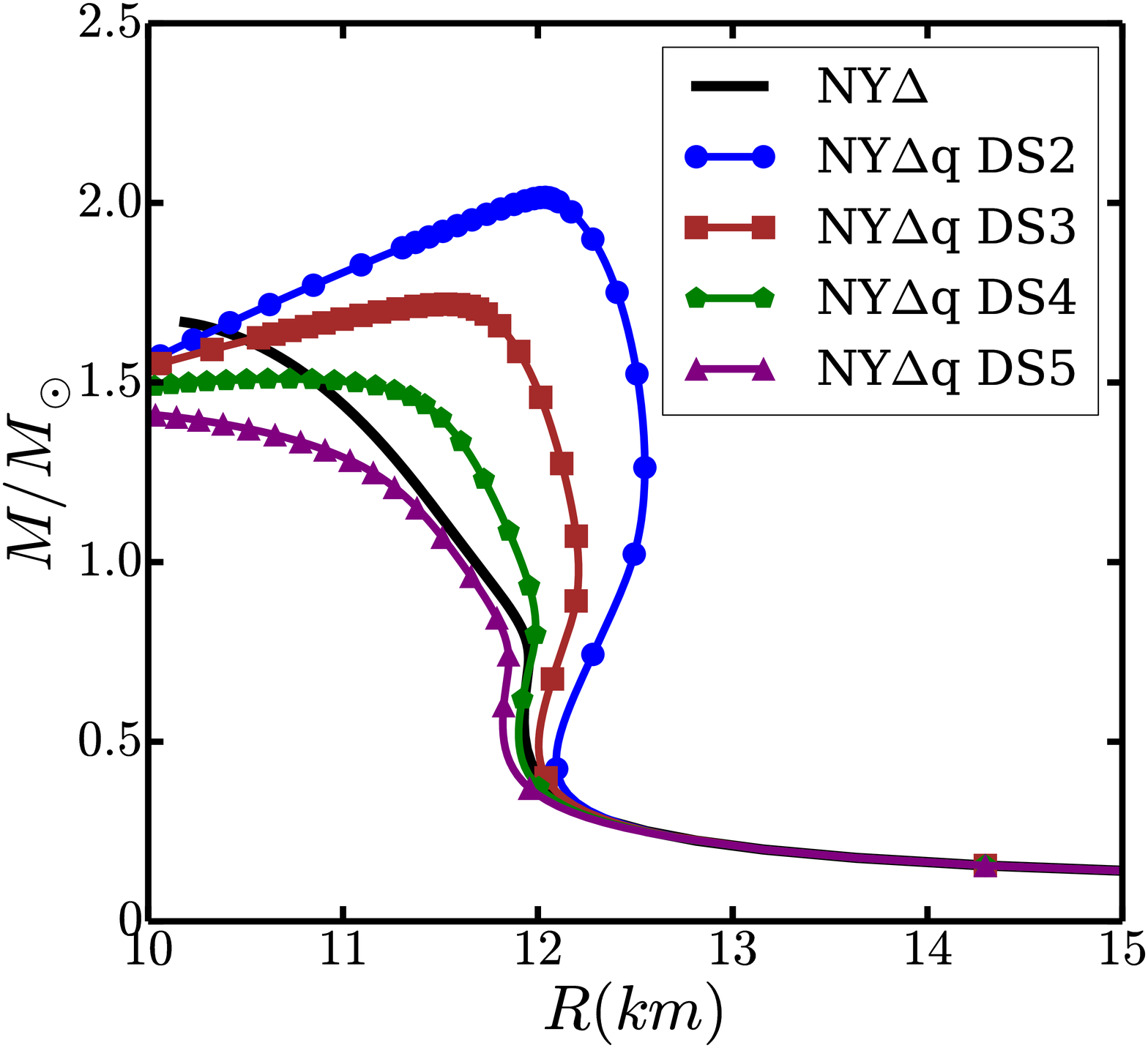}
\caption{The same as Fig.~\ref{fig:mr3Nq} except for that
the solid line corresponds to the result of the neutron star whose ingredients include hyperons and $\Delta$-baryons.}
\label{fig:mr3NYq}
\end{figure}

The obtained results of the mass-radius relation of the hybrid star with the EoS of the matter being constructed by the 3-window interpolation with parameters $(\bar{\rho}, \, \Gamma) = (3.75, \, 1.5)\, \rho_{\textrm{sat}}^{}$ and that of the neutron star whose ingredient matter not including or including hyperons and $\Delta$-baryons are illustrated in Fig.~\ref{fig:mr3Nq},  Fig.~\ref{fig:mr3NYq}, respectively.

One can see from the figures that, just like those in case of the Gibbs construction,
the maximum mass and the corresponding radius of the hybrid star decrease with the increasing of the parameter $\alpha$ in the DSE approach.
However, comparing with the case under the Gibbs construction in which the maximum mass of the hybrid star is generally smaller than that of neutron star,
one can recognize that, by 3-window interpolation construction,
even though the curves of the mass-radius relation differ in shape, the maximum mass can be the same.
Concretely, the maximum mass of the hybrid star without the inclusion of hyperons and with quark parameter $\alpha=2$ is $2.06 \, M_{\odot}^{}$, which is exactly the same as that of pure neutron star.
Nevertheless, the radius of the hybrid star in case of DS2 shown in Fig.~\ref{fig:mr3Nq} is generally larger than that of the pure nucleon star.
The radius of the hybrid star with the maximum mass is $12.40\, \textrm{km}$,
and that of the $1.4\, M_{\odot}^{}$ hybrid star is $12.64\,\textrm{km}$.

As for the results in case with hyperons and $\Delta$-baryons, Fig.~\ref{fig:mr3NYq} displays that,
the inclusion of the quark phase can even increase the maximum mass of the star.
In details, in case of including hyperons and $\Delta$-baryons, the maximum mass of the hadron star is only $1.69\, M_{\odot}^{}$,
but the maximum mass of the hybrid star with quark parameter $\alpha=2$ is $2.01\, M_{\odot}^{}$,
exceeding 2-solar mass.
The radius of the hybrid star including the quark matter in case of DS2, moreover,
is larger than that of the hadron star with the same mass except for the cases that the mass is very small.
The radius corresponding to the maximum mass of the hybrid star is $12.03\, \textrm{km}$,
while that of the $1.4\, M_{\odot}^{}$ hybrid star is $12.53\, \textrm{km}$.

Comparing Fig.~\ref{fig:mr3Nq} and Fig.~\ref{fig:mr3NYq},
one can see that though the mass-radius relation curves of the neutron stars with and without the inclusion of hyperons and $\Delta$-baryons differ greatly,
the curves of the hybrid stars with the same quark parameter remains similar.
This may imply that the appearance of hyperons and $\Delta$-baryons affects the EoS of the hybrid star matter under the 3-window interpolation construction quite slightly.

To summarize the main results of the properties of the stars and for the convenience of further discuss, we list our obtained maximum mass, the corresponding radius and center density of the hybrid star whose EoS of the composing matter is fixed with different construction schemes in Table~\ref{tab:m-r-rhoc}.

\begin{table}[!htp]
\caption{Calculated results of the maximum mass, the corresponding radius and center density of the pure hadron star (NS) and pure quark star (QS), and those of the hybrid stars (HS) whose EoS of the composing (hybrid) matter is determined with different construction schemes and different parameters for the quark sector.}
\label{tab:m-r-rhoc}
\begin{tabular}{|cc|ccc|}
\hline
&&$M_{\textrm{max}}^{}/M_{\odot}^{}$&$R(M_{\textrm{max}})$(km)&$\rho_{c}^{}/\rho_{\textrm{sat}}^{}$ \\[0.5mm]
\hline
          & N                & 2.06 & 10.53 & 7.18 \\
NS      & NY              & 1.69 & 9.93  & 8.76 \\
          & NY$\Delta$ & 1.66 & 10.18 & 7.71 \\
\hline
          & DS2 & 1.27 & 7.66 & 10.59 \\
QS      & DS3 & 1.26 & 7.53 & 11.39 \\
          & DS4 & 1.28 & 7.52 & 11.88 \\
          & DS5 & 1.32 & 7.58 & 11.90 \\
\hline
          & Nq DS2 & 1.89 & 11.05 & 6.81\\
HS      & Nq DS3 & 1.72 & 10.94 & 7.25\\
Gibbs & Nq DS4 & 1.57 & 10.75 & 7.25\\
          & Nq DS5 & 1.47 & 9.96   & 8.80\\
\hline
                 & Nq DS2 & 2.06 & 12.40 & 5.39\\
HS             & Nq DS3 & 1.76 & 12.10 & 5.32\\
3-window  & Nq DS4 & 1.53 & 11.81 & 5.33\\
                 & Nq DS5 & 1.40 & 9.71   & 9.87\\
\hline
                 & NY$\Delta$q DS2 & 2.01 & 12.03 & 5.72 \\
HS             & NY$\Delta$q DS3 & 1.72 & 11.53 & 6.04 \\
3-window  & NY$\Delta$q DS4 & 1.51 & 10.68 & 7.38 \\
                 & NY$\Delta$q DS5 & 1.42 & 9.64  & 9.74 \\
\hline
\end{tabular}
\end{table}

\subsection{Composing Particle Configuration}

One can notice easily from Table~\ref{tab:m-r-rhoc} that the pure nucleon star can be quite massive with a maximum mass exceeding $2\, M_{\odot}^{}$ but either the neutron star whose composing matter includes hyperons or both hyperons and $\Delta$-baryons or the pure quark star can not be so massive. Nevertheless the hybrid star whose EoS of the composing matter is fixed with the 3-window interpolation scheme can have a maximum mass about $2\, M_{\odot}^{}$,
i.e., the 3-window interpolation construction scheme can solve the ``hyperon puzzle" (and the ``$\Delta$ puzzle").
To show this more explicitly and understand the mechanism,
we resort to the composing particle configuration in the hybrid star matter.

\begin{figure}[htb]
\includegraphics[width=0.46\textwidth]{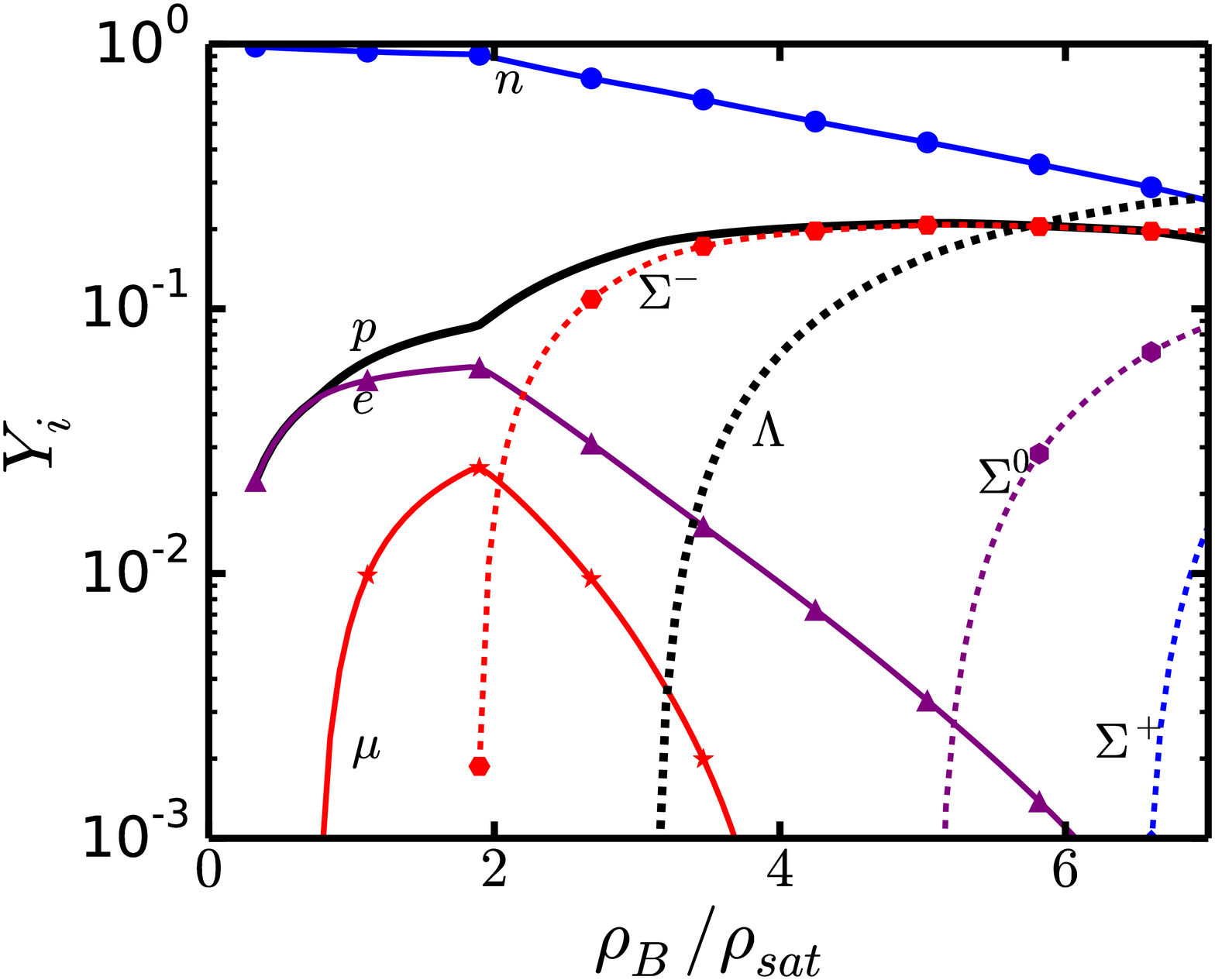}
\caption{Calculated result of the particle fraction of the hadron matter including nucleons and hyperons. }
\label{fig:YiRhoNY}
\end{figure}

\begin{figure}[htb]
\includegraphics[width=0.46\textwidth]{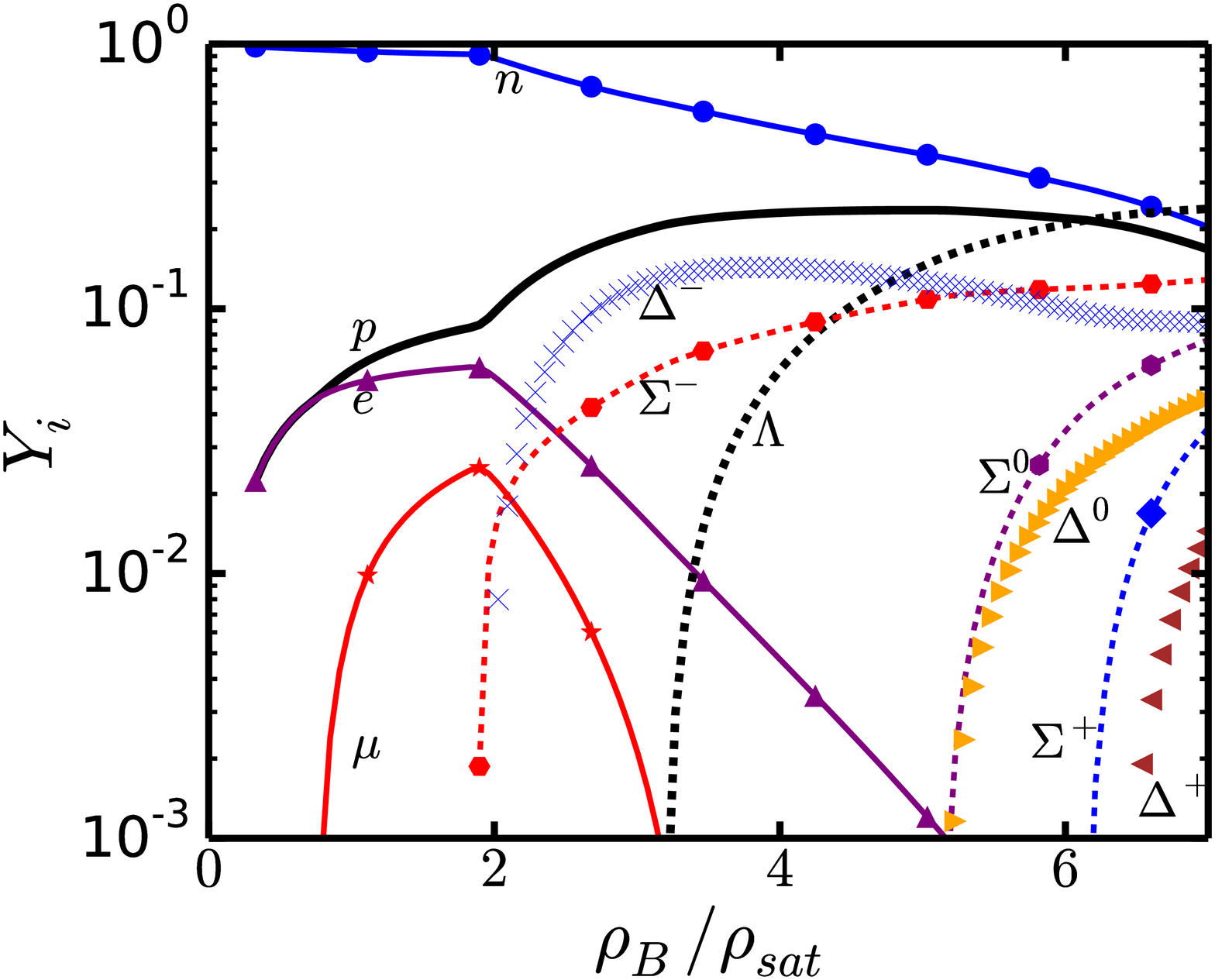}
\caption{Calculated result of the particle fraction of the hadron matter including nucleons, hyperons and $\Delta$-baryons.  }
\label{fig:YiRhoNYD}
\end{figure}

At first we show the calculated results of the particle fraction as a function of baryon density for the hadron matter in Fig.~\ref{fig:YiRhoNY} and Fig.~\ref{fig:YiRhoNYD},
with hyperon only and with both hyperon and $\Delta$-baryon, respectively.
Comparing Figs.~\ref{fig:YiRhoNY} and \ref{fig:YiRhoNYD} one can see that the hyperons begin to appear at 2$\rho_{\textrm{sat}}$, and $\Delta$-baryons emerge also at that density.

One can see further from Figs.~\ref{fig:YiRhoNY} and \ref{fig:YiRhoNYD} that the inclusion of $\Delta$-baryons suppresses the fraction of the $\Sigma^{-}$ hyperon, electron and muon,
and the fraction of neutron is also suppressed at high density.
It means that the $\Delta$-baryons (especially the $\Delta^{-}$) replaces the hyperons (especially the $\Sigma^{-}$). Such a simultaneous appearance of the hyperons and $\Delta$-baryons and the replacement induce that the EoS of the hadron matter including both hyperons and $\Delta$-baryons is almost the same as that not including $\Delta$-baryons, just as the Figs.~\ref{fig:p-muB} and \ref{fig:p-e} have shown. In turn the ``$\Delta$ puzzle" and the ``hyperon puzzle" appear simultaneously.

\begin{figure}[htb]
\includegraphics[width=0.48\textwidth]{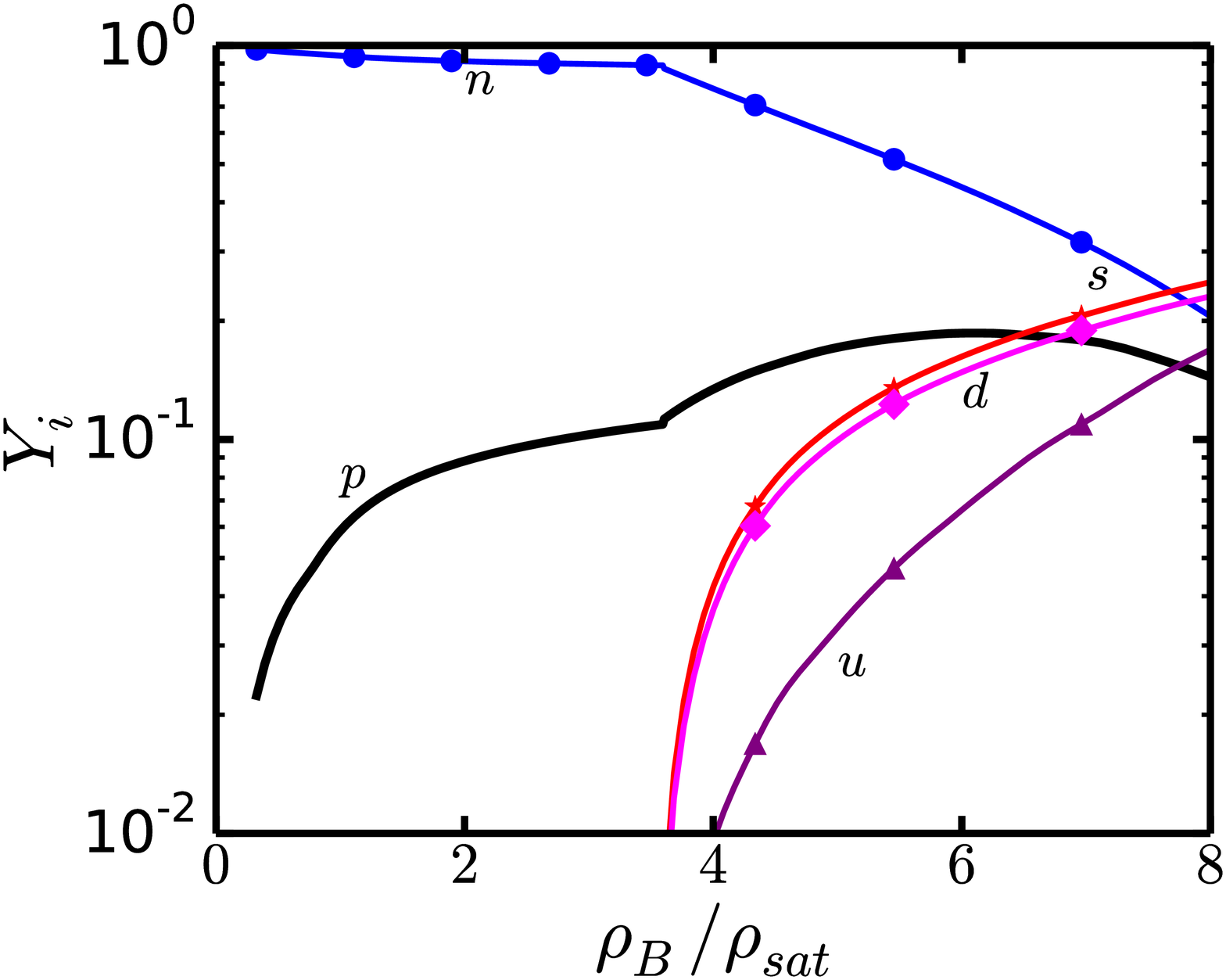}
\caption{Calculated result of the baryon density dependence of the particle fraction in the hybrid star matter under the Gibbs construction, where the hadron matter does not include either hyperons or   $\Delta$-baryons, and the parameter describing the property of the quark matter is $\alpha=2$.   $Y_{i}^{}={\rho_{i}^{}}/{\rho_{B}^{}}$ for the hadron sector and $Y_{i}^{}={\rho_{i}^{}}/3{\rho_{B}^{}}$ for the quark sector.}
\label{fig:YiRhoGNq}
\end{figure}

\begin{figure}[htb]
\includegraphics[width=0.48\textwidth]{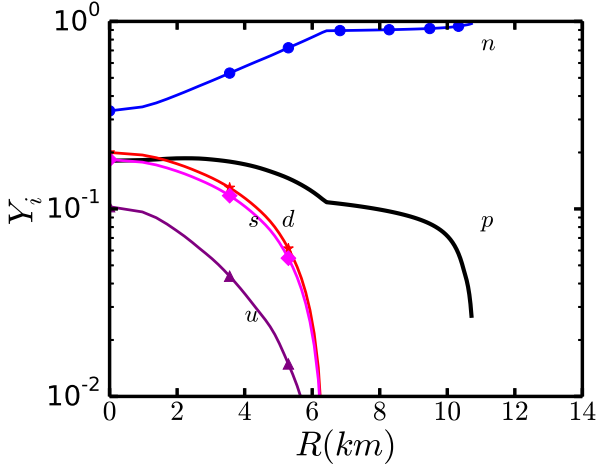}
\caption{Calculated variation behavior of the particle fraction of the matter in terms of the distance from the center of the maximal mass hybrid star under the Gibbs construction.
The other items are the same as Fig.~\ref{fig:YiRhoGNq} . }
\label{fig:YiRGNq}
\end{figure}

The calculated result of the baryon density dependence of the particle fraction of the hybrid matter under the Gibbs construction and the variation behavior of the particle fraction of the matter in terms of the distance from the center of the maximum mass hybrid star are illustrated in Fig.~\ref{fig:YiRhoGNq},
Fig.~\ref{fig:YiRGNq}, respectively. The quark sector in the Gibbs construction is that described with the  parameter $\alpha=2$.
The two figures manifest apparently that in the low density region (the outer part of the star) the composing particles of the matter are pure hadrons, and the quarks appear as the density is about 3--$4\,\rho_{\textrm{sat}}^{}$ (exist in the core with a radius about $6\,\textrm{km}$).
And at about $8\rho_{\textrm{sat}}^{}$, which is beyond the center density of the maximum mass hybrid star, the number fraction of the quarks and hadrons are of the same order.
Since the lower density hadron mantle region (composed of nucleons) is quite large (more than 5 times the volume of the hybrid matter region) and the EoS of the hadron matter at low baryon density is rather soft (as shown in Figs.~\ref{fig:p-e}, \ref{fig:eosGNq}, \ref{fig:rhobar-Gamma} and \ref{fig:eos3Nq} ) and that of the hybrid matter is much softer (see Fig.~\ref{fig:eosGNq}), such a hybrid star can then not be very massive.

\begin{figure}[htb]
\includegraphics[width=0.45\textwidth]{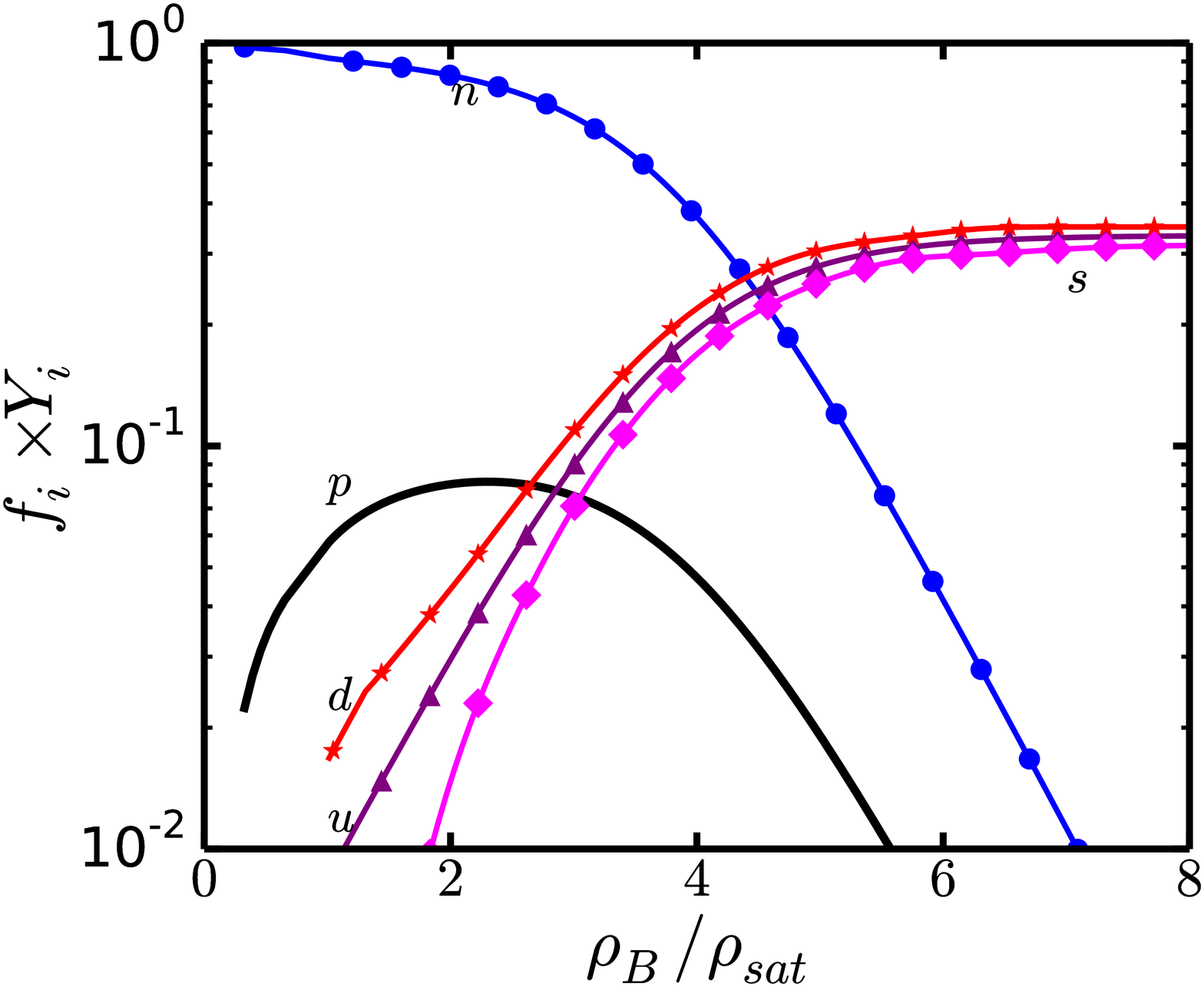}
\caption{Calculated result of the baryon density dependence of the particle fraction in the hybrid star matter under the 3-window interpolation construction, where the hadron matter does not include either the hyperons or the  $\Delta$-baryons, and the parameter to describe the property of the quark matter is $\alpha=2$.
 $Y_{i}^{}={\rho_{i}^{}}/{\rho_{B}^{}}$ for the hadron sector and $Y_{i}^{}={\rho_{i}^{}}/3{\rho_{B}^{}}$ for the quark sector,  and $f_{i}^{} = f_{H,Q}^{} = f_{-,+}^{}$ is that defined in Eq.~(\ref{eqn:fHfQ}).}
\label{fig:YiRho-3Nq}
\end{figure}

\begin{figure}[htb]
\includegraphics[width=0.45\textwidth]{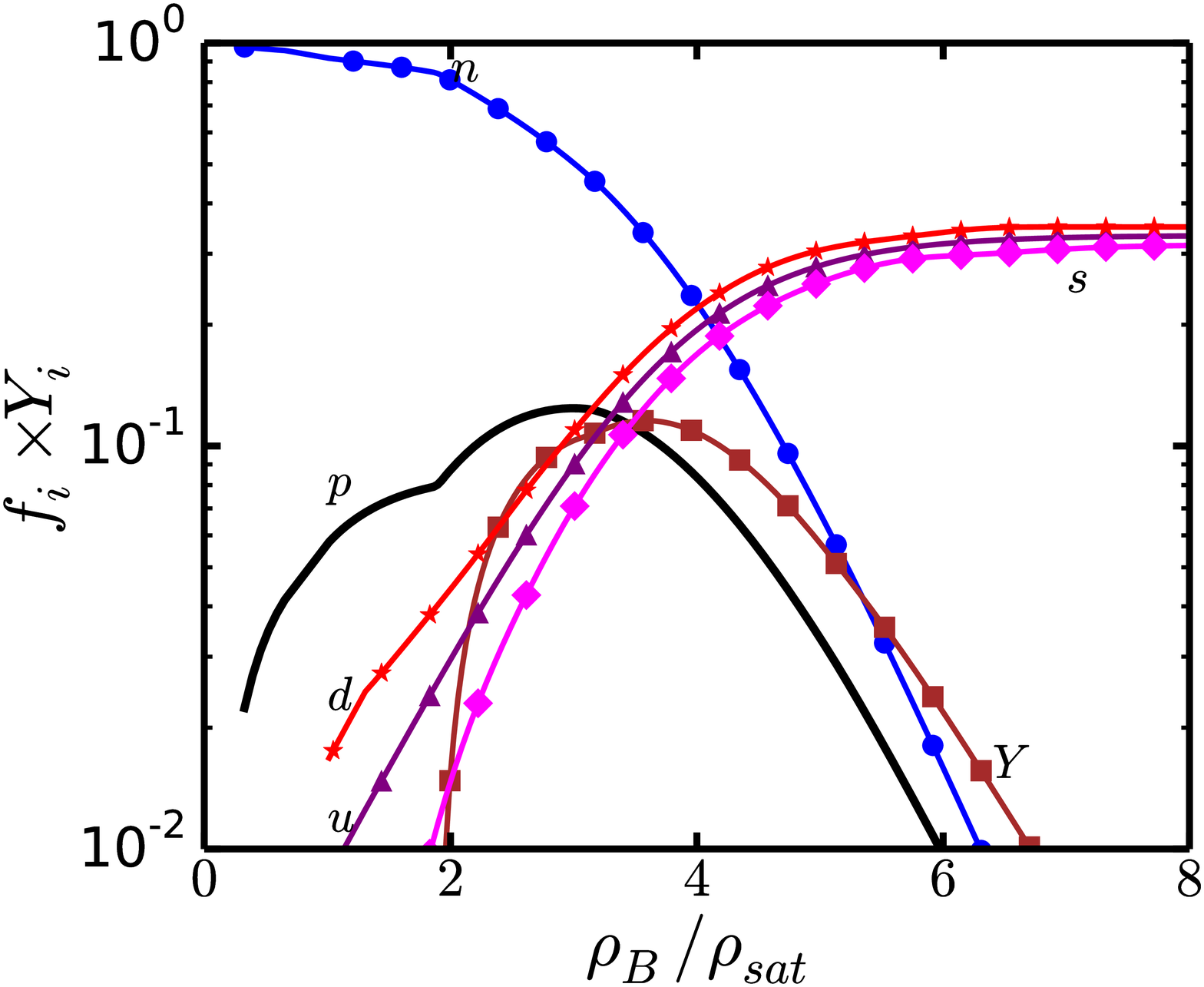}
\caption{The same as Fig.~\ref{fig:YiRho-3Nq}, but the hadron matter sector in the construction including hyperons.}
\label{fig:YiRho3NYq}
\end{figure}

\begin{figure}[htb]
\includegraphics[width=0.45\textwidth]{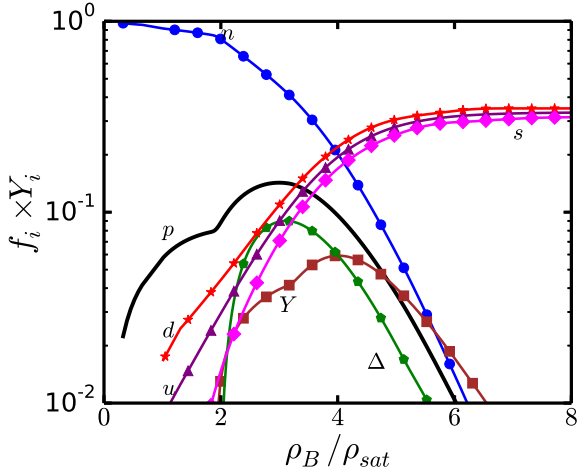}
\caption{The same as Fig.~\ref{fig:YiRho-3Nq}, but the hadron matter sector in the construction including both hyperons and $\Delta$-baryons.}
\label{fig:YiRho3NYDq}
\end{figure}

\begin{figure}[htb]
\includegraphics[width=0.46\textwidth]{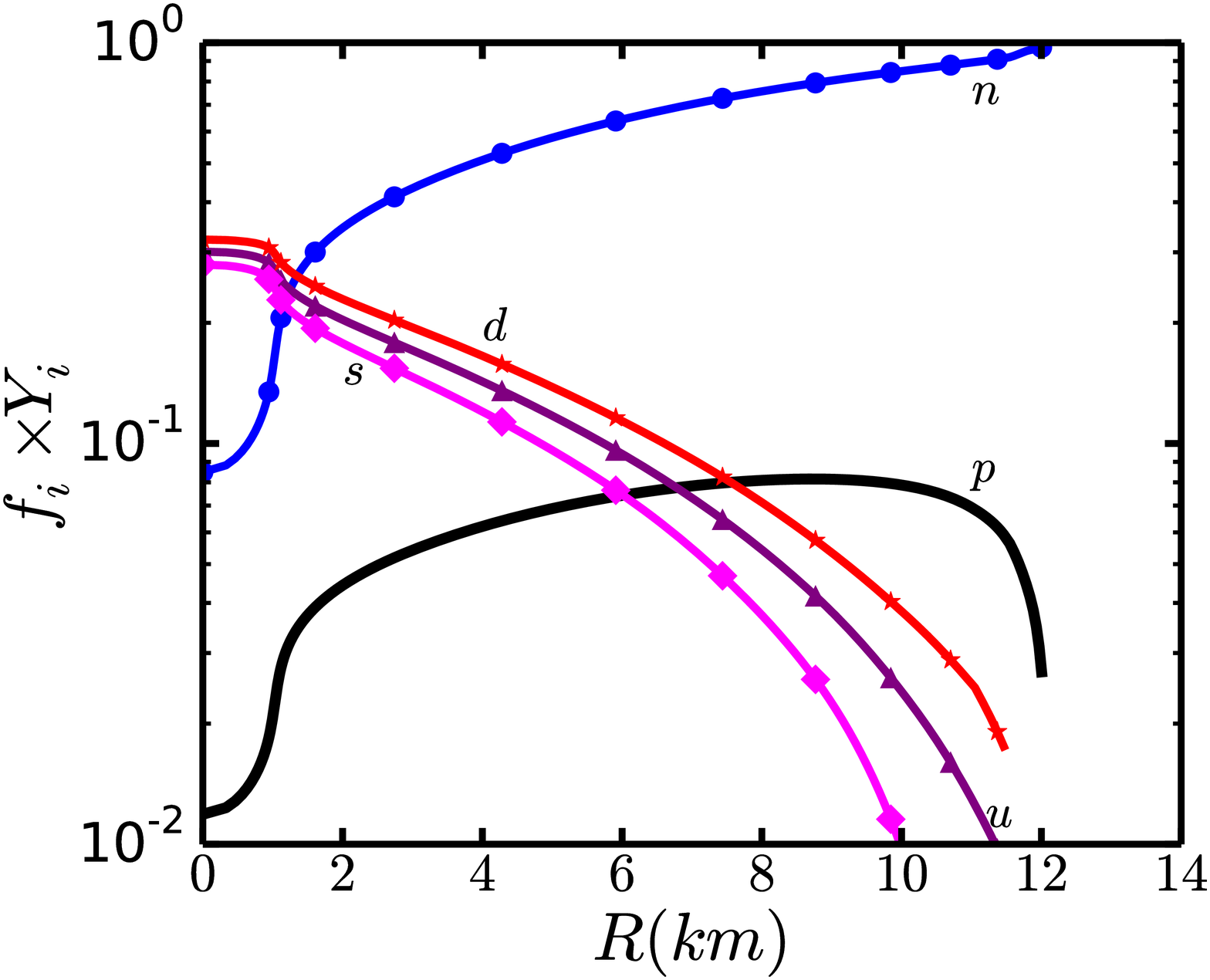}
\caption{Calculated variation behavior of the particle fraction of the matter in terms of the distance from the center of the maximum mass hybrid star under the 3-window interpolation construction.
The hadron phase in the hybrid matter does not include hyperons and $\Delta$-baryons.
The parameter describing the property of the quark sector is $\alpha=2$.
 $Y_i=\rho_i/\rho_B$ for the hadron sector and $Y_i=\rho_i/3\rho_B$ for the quark sector,
 and $f_i$ is that defined in Eq.~(\ref{eqn:fHfQ}).}
\label{fig:YiR-3Nq}
\end{figure}

\begin{figure}[htb]
\includegraphics[width=0.46\textwidth]{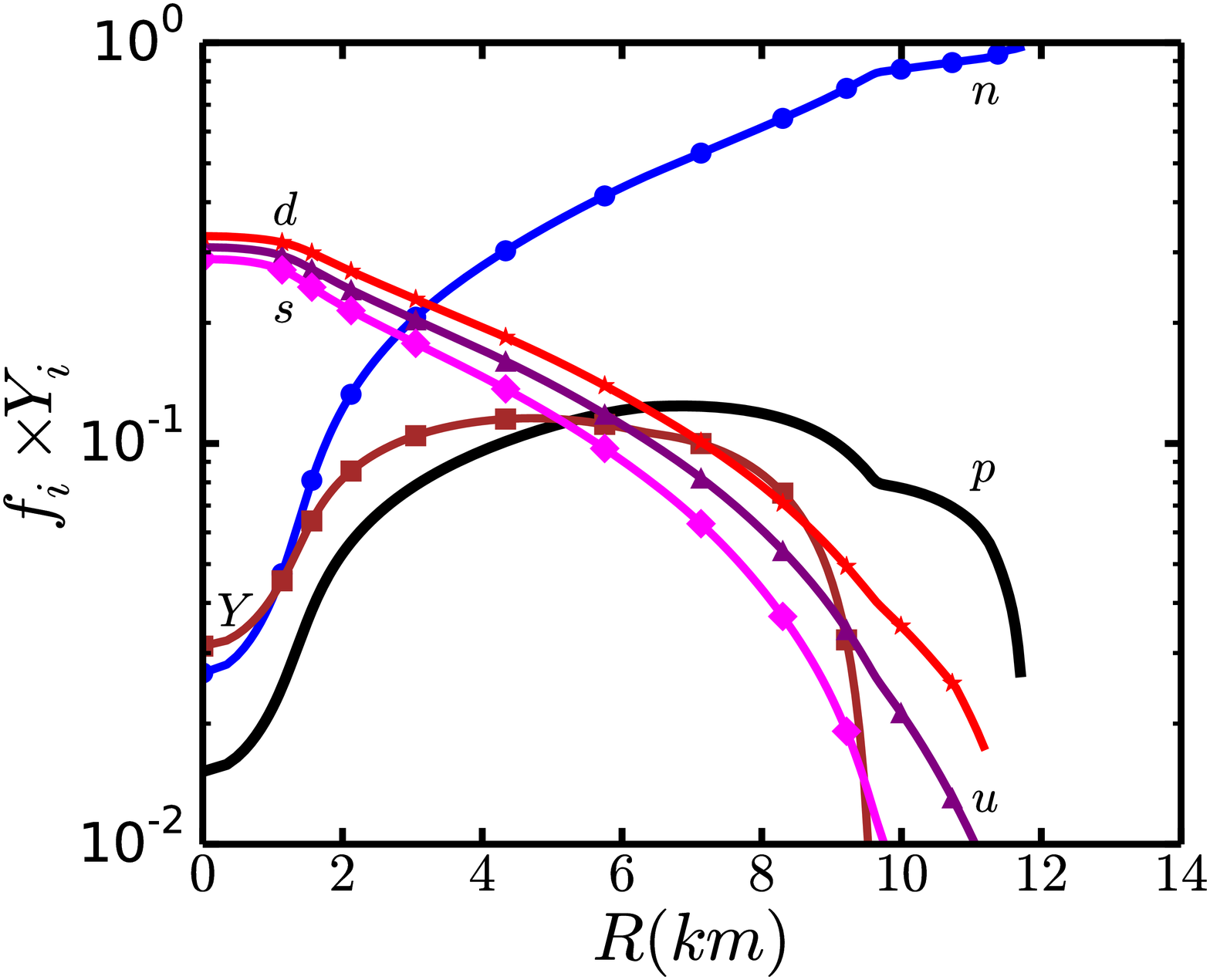}
\caption{The same as Fig.~\ref{fig:YiR-3Nq}, but for that the hadron matter sector in the construction includes hyperons.}
\label{fig:YiR3NYq}
\end{figure}

\begin{figure}[htb]
\includegraphics[width=0.46\textwidth]{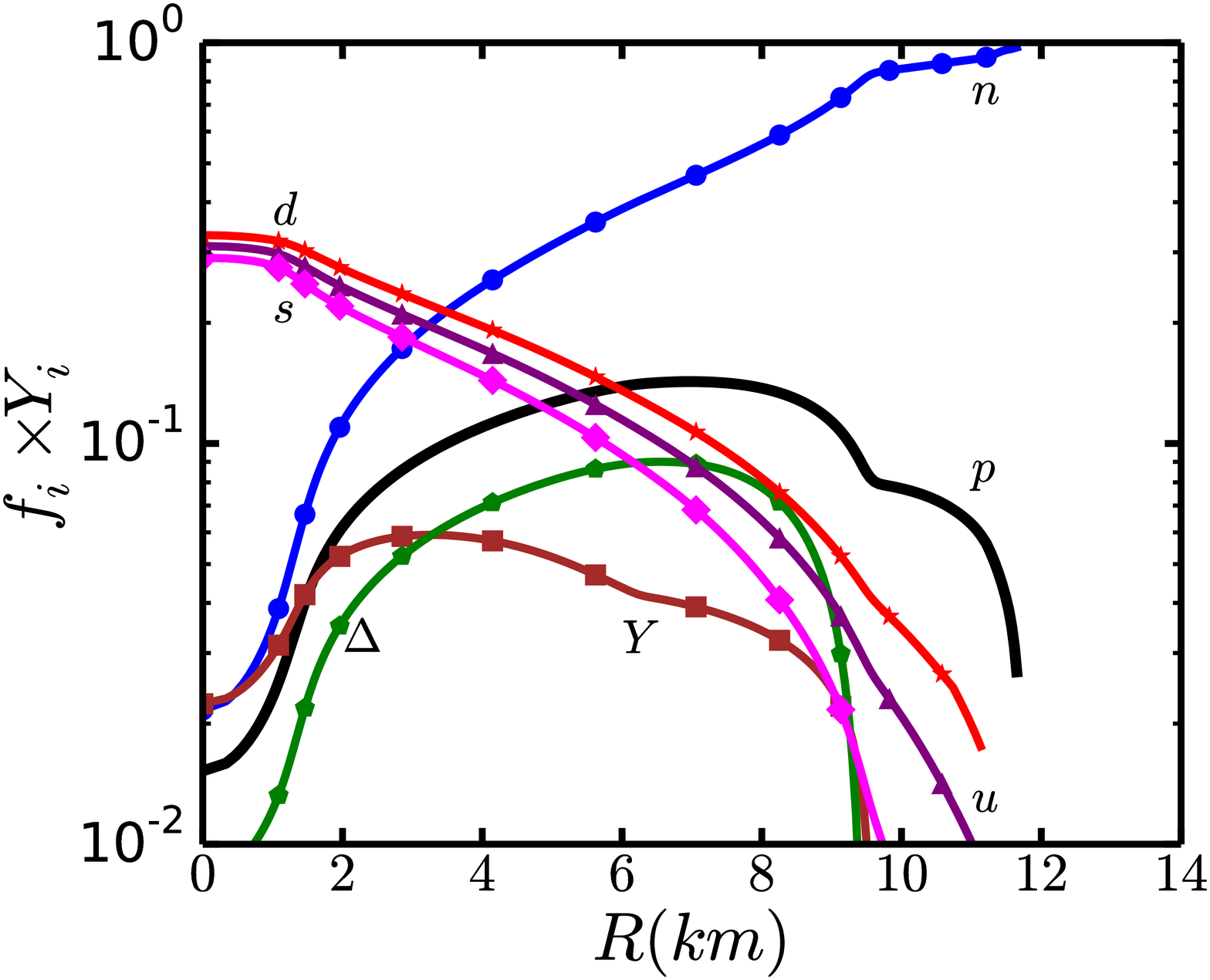}
\caption{The same as Fig.~\ref{fig:YiR-3Nq}, but for that the hadron matter sector in the construction includes both hyperons and $\Delta$-baryons. }
\label{fig:YiR3NYDq}
\end{figure}

We have also calculated the baryon density dependence of the particle fraction of the hybrid matter under the 3-window interpolation construction, as well as the variation behavior of the particle fraction of the matter in terms of the distance from the center of the maximum mass hybrid star.
The obtained result of the baryon density dependence of the particle fraction in the hybrid matter whose  hadron sector consists of only nucleons is shown in Fig.~\ref{fig:YiRho-3Nq}.
Those for the cases that the hadron matter sector includes hyperons, both hyperons and $\Delta$-baryons are displayed in Fig.~\ref{fig:YiRho3NYq}, Fig.~\ref{fig:YiRho3NYDq}, respectively.
The calculated results of the variation behavior of the particle fraction of the matter in terms of the distance from the center of the maximum mass hybrid star are illustrated in Fig.~\ref{fig:YiR-3Nq}, Fig.~\ref{fig:YiR3NYq},
Fig.~\ref{fig:YiR3NYDq} for the three cases of the hadron matter, respectively.
The quark sector in the 3-window interpolation construction is also that described with the  parameter $\alpha=2$.
Since we are using the interpolation of the energy contribution to describe the phase transition region,
the direct particle fraction $Y_{i}^{}$ does not have the usual meaning.
Therefore, we take $f_{i}^{} \times Y_{i}^{} $ (with $f_{i}^{}$ defined in Eq.~(\ref{eqn:fHfQ}))
to identify the particle fraction, which represents the effect of a certain species of the particles on the EoS.

From Figs.~\ref{fig:YiRho-3Nq} -- \ref{fig:YiR3NYDq}, one can recognize distinctly that the quark matter appears at very low baryon density (according to our interpolation scheme, the contribution factor
$f_{Q}^{} = f_{+}^{} \sim 0.01$ at $\rho_{B}^{} = \rho_{\textrm{sat}}$).
The hyperons and $\Delta$-resonances appear as $\rho_{B}^{} \approx 2\rho_{\textrm{sat}}^{}$,
or in other word, exist in the region $R < 10\,\textrm{km}$ (in more detail, the $\Delta$-resonances
can not exist in the region very close to the center).
These facts manifest that the hadron mantle of the hybrid star consists of only nucleons and is very thin,
the hybrid star is thus composed mainly of the hybrid matter.
Referring to Figs.~\ref{fig:eos3Nq}, \ref{fig:eos3NYDq}, \ref{fig:pnB3Nq} and \ref{fig:pnB3NYDq},
one can know that the EoS of the hybrid star matter under the 3-window interpolation construction
(no matter the hadron matter sector includes hyperons and $\Delta$-baryons or not) is quite stiff (even much stiffer than that of the corresponding hadron matter's in the $\rho_{B}^{} \in (2$--$4) \rho_{\textrm{sat}}^{}$ region).
As a consequence, the maximum mass of the hybrid star can be as massive as exceeding $2\, M_{\odot}^{}$, almost the same as that of the pure nucleon star.

It is also remarkable that as the compact star with maximum mass exceeding $2\, M_{\odot}^{}$ is obtained (or, in other word, the ``hyperon puzzle" and the ``$\Delta$ puzzle" are solved)
when one takes the quark degrees of freedom (or the hadron-quark phase transition) into account, there does not exist pure massive quark star, even there is no pure quark core inside the maximum mass hybrid star.
In Gibbs construction for the hybrid star matter, the core consists of considerable fraction (about $51\%$) of nucleons.
While in the 3-window interpolation scheme, quark matter contributes definitely most (about $94\%$) to the center region of the star but there still exists nucleons and hyperons (about $6\%$), and considerable  amount of hyperons and $\Delta$-resonances appears in the middle region (about $13\%$).
The mechanism for the solving of the so called ``hyperon puzzle" and the ``$\Delta$ puzzle" can then be attributed to that the hadron-quark phase transition (the mixing of the hadron matter and the quark matter) stiffens the EoS of the hybrid matter at middle and high density region.

\section{\label{sec:sum}Summary and Remarks}

We have investigated the mass-radius relation of hybrid stars with both the Gibbs construction and the 3-window interpolation construction for the EoSs of the composing matters in this paper.
For that of hadron phase we adopt the result of the relativistic mean field theory,
and for that of quark phase we take the result via the Dyson-Schwinger equation approach of QCD.

Our calculation manifests that the Gibbs construction results in a rather soft EoS for the hybrid star matter,
while for the 3-window interpolation,
the EoS of the hybrid star matter can be stiffer than those in both the hadron phase and the quark phase separately.
Therefore, for a hybrid star whose hadron matter sector includes hyperons and $\Delta$-resonances,
the maximum mass can exceed $2\,M_{\odot}^{}$, which is in accordance with the observations
several years ago as well as the upper limitation of maximum mass of $\sim2.17M_{\odot}$ by recent gravitational wave estimation~\cite{Margalit:2017}.

It indicates that taking the hadron-quark phase transition into account with the 3-window interpolation scheme to construct the EoS of the hybrid star matter can solve the ``hyperon puzzle" and the ``$\Delta$ puzzle".
 Nevertheless the high-mass compact star is not pure quark star, even not hybrid star with a pure quark core.
More concretely, the matter around the center involves nucleons and hyperons, and that in the middle region even has $\Delta$-baryons.
This provides a further evidence for the happening of the hadron-quark phase transition in the core of the compact star and the significance of the phase coexistence in governing the properties of compact stars.

For the radius of the compact stars,
our result of the hybrid star seems larger than some of the estimates from observational data at first glance.
For example, the radius of the neutron star with canonical mass, $1.4M_\odot$,
is estimated to be $R_{1.4} \in (9.7, \, 13.7)\,$km based on chiral effective theory~\cite{Hebeler:2010PRL},
or $9.4\pm 1.2\,$km by analysing the quiescent x-ray transients in low mass x-ray binaries~\cite{Guillot:2014ApJL}.
However, our present result coincides with some of the previous results (see, e.g. Refs.~\cite{Oertel:2015JPG,Dutra:2016PRC}),
and also the most recent estimation via gravitational wave~\cite{Bauswein:2017}.
Furthermore, in view of the result that the effective radius of a hadron increases with respect to the
increase of the density (or temperature) of the strong interaction matter (see, e.g. Refs.~\cite{Liu:20013NPA,Chang:2005NPA,Liu:2010PRC,Wang:2013PRD}),
we believe our present result is reasonable.

Analyzing the  detail of our calculation and the obtained results,
we confirm that the maximum mass of the neutron star is determined mainly by the stiffness of the EoS at density above $ (2\sim 4)\rho_{\textrm{sat}}^{}$,
and the slope of the mass-radius relation curve is related closely with the EoS in the range of
$\rho_{B}^{} \approx (2 \sim 4) \rho_{\textrm{sat}}^{}$~\cite{Lattimer:2001ApJ,Lattimer:200716PRs,Ozel:2009PRD}.
In turn, an EoS which is stiff at middle and high densities while soft at lower density can be expected as the perfect one.
And this may be reached by changing the parameters describing the quark phase.
Recalling our calculation process, we know that the parameter $\alpha$ in Eq.~(\ref{eqn:alpha}) in the DSE approach affects the stiffness of the EoS in quark phase at densities relevant that in the core of the star,
and the bag constant in Eq.~(\ref{eqn:B_DS}) determines the pressure at zero density.
Therefore, it is possible to get a perfect EoS by adjusting the $\alpha$ and $B_{\textrm{DS}}$.
The relevant work is under progress.

\begin{acknowledgments}
The work was supported by the National Natural Science Foundation of China under Contract No.~11435001, No.~11305144 and No.~11475149; the National Key Basic Research Program of China under Contracts No.~G2013CB834400 and No.~2015CB856900.
\end{acknowledgments}

%\bibliography{refspec}% Produces the bibliography via BibTeX.
%
%\bibliography{ms}% Produces the bibliography via BibTeX.

\begin{thebibliography}{120}

\bibitem{Shapiro:1983}
      S.L. Shapiro, and S.A. Teukolsky,
	    {\it Black Holes, White Dwarfs, and Neutron Stars} (John Wiley \& Aons, New York, 1983)

\bibitem{Glendenning:2000}
      N. K. Glendenning,
         {\it Compact Stars: Nuclear Physics, Particle Physics, and General Relativity}
            (Springer-Verlag, Berlin, 2000).

\bibitem{Castillo:2016EPJA}
      D. Alvarez-Castillo, S. Benic, D. Blaschke, S. Han, and S. Typel,
         Eur. Phys. J. A {\bf 52}, 232 (2016).
% ``  Neutron star mass limit at 2 M-circle dot supports the existence of a CEP. "

\bibitem{Lattimer:2001ApJ}
      J.M. Lattimer, and M. Prakash,
         Astrophys. J. {\bf 550}, 426 (2001).
% ``  Neutron star structure and the equation of state."

\bibitem{Lattimer:200716PRs}
      J.M. Lattimer, and M. Prakash,
         Phys. Rept. {\bf 442}, 109 (2007);
% ``  Neutron star observations: Prognosis for equation of state constraints."
         Phys. Rept. {\bf 621}, 127 (2016).
% `` The equation of state of hot, dense matter and neutron stars."

\bibitem{Ozel:2009PRD}
      F. Ozel, and D. Psaltis,
         Phys. Rev. D {\bf 80}, 103003 (2009).
%  `` Reconstructing the neutron-star equation of state from astrophysical measurements."

\bibitem{Oertel:2017RMP}
       M. Oertel, M. Hempel, T. Klahn, S. Typel,
          Rev. Mod. Phys. {\bf 89}, 015007 (2017).
% `` Equations of state for supernovae and compact stars"

\bibitem{Baldo:1997AA}
     M. Baldo, I. Bombaci, and G.F. Burgio,
        Astron. Astrophys. {\bf 328}, 274 (1997).
% ``  Microscopic nuclear equation of state with three-body forces and neutron star structure. "

\bibitem{Akmal:1998PRC}
      A. Akmal, V.R. Pandharipande, and D.G. Ravenhall,
         Phys. Rev. C {\bf 58}, 1804 (1998).
% `` Equation of state of nucleon matter and neutron star structure."

\bibitem{Zhou:2004PRC}
      X.R. Zhou, G.F. Burgio, U. Lombardo, H.-J. Schulze, and W. Zuo,
         Phys. Rev. C {\bf 69}, 018801 (2004).
% `` Three-body forces and neutron star structure."

\bibitem{Li:2008PRC}
     Z.H. Li, and H.-J. Schulze,
        Phys. Rev. C {\bf 78}, 028801 (2008).
% `` Neutron star structure with modern nucleonic three-body forces."

\bibitem{Bunta:2004PRC}
      J. K. Bunta, and S. Gmuca,
         Phys. Rev. C {\bf 70}, 054309 (2004).
% `` Hyperons in a relativistic mean-field approach to asymmetric nuclear matter."

\bibitem{Schulze:2006PRC}
      H.-J. Schulze, A. Polls, A. Ramos, and I. Vidana,
          Phys. Rev. C {\bf 73}, 058801 (2006).
% `` Maximum mass of neutron stars."

\bibitem{Sedrakian:2007PPNP}
      A. Sedrakian,
         Prog. Part. Nucl. Phys. {\bf 58}, 168 (2007).
% ``  The physics of dense hadronic matter and compact stars."

\bibitem{Shao:2009PRC}
      G.Y. Shao, and Y.X. Liu,
         Phys. Rev. C {\bf 79}, 025804 (2009).
% `` Influence of the sigma-omega meson interaction on neutron star matter."

\bibitem{Shao:2010PRC}
      G.Y. Shao, and Y.X. Liu,
         Phys. Rev. C {\bf 82}, 055801 (2010).
% `` Influence of the isovector-scalar channel interaction on neutron star matter with hyperons and
%    antikaon condensation ."

\bibitem{Fortin:2017PRC}
      M. Fortin, S. S. Avancini, C. Provid\^encia, and I. Vida\~na,
         Phys. Rev. C {\bf 95}, 065803 (2017).
% `` Hypernuclei and massive neutron stars."

\bibitem{Demorest:2010Nature}
      P.B. Demorest, T. Pennucci, S.M. Ransom, M.S.E. Roberts, and J.W.T. Hessels,
         Nature {\bf 467}, 1081 (2010).
% `` A two-solar-mass neutron star measured using Shapiro delay."

\bibitem{Antoniadis:2013Science}
      J. Antoniadis {\it et al.},
          Science {\bf 340}, 1233232 (2013).
% `` A Massive Pulsar in a Compact Relativistic Binary."

\bibitem{Lonardoni:2015PRL}
      D. Lonardoni, A. Lovato, S. Gandolfi, and F. Pederiva,
         Phys. Rev. Lett {\bf 114}, 092301 (2015).
% ``  Hyperon Puzzle: Hints from Quantum Monte Carlo Calculations. "

\bibitem{Masuda:2016EPJA}
      K. Masuda, T. Hatsuda and T. Takatsuka,
         Eur. Phys. J. A {\bf 52}, 65 (2016).
% `` Hyperon puzzle, hadron-quark crossover and massive neutron stars. "

\bibitem{Hu:2003PRC}
      X. Hu, and H. Guo,
         Phys. Rev. C {\bf 67},038801 (2003).
%``Delta excitation and its influences on neutron stars in relativistic mean field theory''

\bibitem{Schurhoff:2010ApJ}
      T. Sch\"urhoff, S. Schramm, and V. Dexheimer,
          Astrophys. J {\bf 724}, L74 (2010).
%``Neutron stars with small radii  The role of Delta resonances''

\bibitem{Lavagno:2010PRC}
      A. Lavagno,
          Phys. Rev. C {\bf 81}, 044909 (2010).
%``Hot and dense hadronic matter in an effective mean-field approach''

\bibitem{Drago:2014PRC}
	A. Drago, A. Lavagno, G. Pagliara, and D. Pigato,
		Phys. Rev. C {\bf 90}, 065809 (2014).
%``Early appearance of Delta isobars in neutron stars''

\bibitem{Cai:2015PRC}
	B. J. Cai, F. J. Fattoyev, B. A. Li, and W. G. Newton,
		Phys. Rev. C {\bf 92}, 015802 (2015).
%``Critical density and impact of Delta(1232) resonance formation in neutron stars''

\bibitem{Drago:2016EPJA}
	A. Drago, A. Lavagno, G. Pagliara and D. Pigato,
         Eur. Phys. J. A {\bf 52}, 40 (2016).

\bibitem{Lopes:2014PRC}
      L.L. Lopes and D.P. Menezes,
         Phys. Rev. C {\bf 89}, 025805 (2014).
% ``  Hypernuclear matter in a complete SU(3) symmetry group."

\bibitem{Gomes:2015ApJ}
      R.O. Gomes, V. Dexheimer, S. Schramm, and C.A.Z. Vasconcellos,
         Astrophys. J {\bf 808}, 8 (2015).
% `` Many-body Forces in the Equation of State of Hyperonic Matter."

\bibitem{Oertel:2015JPG}
     M. Oertel, C. Provid\^encia, F. Gulminelli, and A. R. Raduta,
        J. Phys. G {\bf 42}, 075202 (2015).
% ``  Hyperons in neutron star matter within relativistic mean-field models."

\bibitem{Katayama:2015PLB}
      T. Katayama, and K. Saito,
          Phys. Lett. B {\bf 747}, 43 (2015).
% ``  Hyperons in neutron stars."

\bibitem{Lonardoni:2013PRC}
      D. Lonardoni, S. Gandolfi, and F. Pederiva,
          Phys. Rev. C {\bf 87}, 041303(R) (2013).
% ``  Effects of the two-body and three-body hyperon-nucleon interactions in Lambda hypernuclei ."

\bibitem{Lonardoni:2014PRC}
      D. Lonardoni, F. Pederiva, and S. Gandolfi,
         Phys. Rev. C {\bf 89}, 014314 (2014).
% `` Accurate determination of the interaction between A hyperons and nucleons from
%   auxiliary field diffusion Monte Carlo calculations."

\bibitem{Chodos:1974PRD}
	A. Chodos, R. L. Jaffe, C. B. Thorn, and V. F. Weisskopf,
		Phys. Rev. D {\bf 9}, 3471 (1974).
% ``  New Extended Model of Hadrons. "

\bibitem{Farhi:1984PRD}
	  E. Farhi, and R. L. Jaffe,
		 Phys. Rev. D {\bf 30}, 2379 (1984).
% `` Strange Matter."

\bibitem{Burgio:2002PRC}
	  G. F. Burgio, M. Baldo, P. K. Sahu, and H.-J. Schulze,
		 Phys. Rev. C {\bf 66}, 025802 (2002).
% `` Hadron-quark phase transition in dense matter and neutron stars . "
%
\bibitem{Nicotra:2006PRD}
	  O. E. Nicotra, M. Baldo, G. F. Burgio, and H.-J. Schulze,
		 Phys. Rev. D {\bf 74}, 123001 (2006).
% ``  Hybrid protoneutron stars with the MIT bag model  "
%
\bibitem{Buballa:1996NPA}
      M. Buballa,
         Nucl. Phys. A {\bf 611}, 393 (1996).
% `` The problem of matter stability in the Nambu-Jona-Lasinio model."

\bibitem{Schertler:1999}
      K. Schertler, S. Leupold, and J. Schaffner-Bielich,
         Phys. Rev. C {\bf 60}, 025801 (1999).
% `` Neutron stars and quark phases in the Nambu-Jona-Lasinio model."

\bibitem{Shao:2011PRDa}
      G.Y. Shao, M. Di Toro, B. Liu, M. Colonna, V. Greco, Y.X. Liu, and S. Plumari,
         Phys. Rev. D {\bf 83}, 094033 (2011).
%`` Hadron-quark phase transition in asymmetric matter with dynamical quark masses."
%
\bibitem{Shao:2011PRDb}
      G.Y. Shao, M. Di Toro, V. Greco, M. Colonna, S. Plumari, B. Liu, and Y.X. Liu,
         Phys. Rev. D {\bf 84}, 034028 (2011).
% `` Phase diagrams in the hadron-Polyakov-Nambu-Jona-Lasinio model."
%
\bibitem{Shao:2013PRD}
      G.Y. Shao, M. Colonna, M. Di Toro, Y.X. Liu, and B. Liu,
         Phys. Rev. D {\bf 87}, 096012 (2013).
% `` Isoscalar-vector interaction and hybrid quark core in massive neutron stars."
%
\bibitem{Klahn:2013PRD}
      T. Kl\"ahn, R. Lastowiecki, and D.Blaschke,
         Phys. Rev. D {\bf 88}, 085001 (2013).
% ``  Implications of the measurement of pulsars with two solar masses for quark matter in compact stars
%   and heavy-ion collisions: A Nambu-Jona-Lasinio model case study ."

\bibitem{Benic:2015AA}
      S. Beni, D. Blaschke, D.E. Alvarez-Castillo, T. Fischer, and S. Typel,
         Astron. Astrophys. {\bf 577}, A40 (2015).
% `` A new quark-hadron hybrid equation of state for astrophysics I. High-mass twin compact stars ."

\bibitem{Chakrabarty:1991PRD}
      S. Chakrabarty,
         Phys. Rev. D {\bf 43}, 627 (1991).
% ``  Equation of State of Strange Quark Matter and strange Star."

\bibitem{Benvenuto:1995PRD}
      O.G. Benvenuto, and G. Lugones,
          Phys. Rev. D {\bf 51}, 1989 (1995).
% `` Strange Matter Equation of State in the Quark Mass-density-dependent Model."

\bibitem{Torres:2013EPL}
      J.R. Torres, and D.P. Menezes,
          Eur. Phys. Lett. {\bf 101}, 42003 (2013).
% `` Quark matter equation of state and stellar properties."

\bibitem{Zacchi:2015PRD}
      A. Zacchi, R. Stiele, and J. Schaffner-Bielich,
         Phys. Rev. D {\bf 92}, 045022 (2015).
% ``  Compact stars in a SU(3) quark-meson model ."

\bibitem{Peshier:2000PRC}
      A. Peshier, B. K\"ampfer, and G. Soff,
         Phys. Rev. C {\bf 61}, 045203 (2000).
% `` Equation of state of deconfined matter at finite chemical potential in a quasiparticle description."

\bibitem{Tian:2012PRD}
      Y.L. Tian, Y. Yan, H. Li, X.L. Luo, and H.S. Zong,
         Phys. Rev. D {\bf 85}, 045009 (2012).
% `` Equation of state of a quasiparticle model at finite chemical potential and quark star."

\bibitem{Zhao:2015PRD}
      T. Zhao, Y. Yan, X.L. Luo, and H.S. Zong,
         Phys. Rev. D {\bf 91}, 034018 (2015).
% `` Study of rotational quark stars and hybrid stars based on the latest equation of state
%   and observation data."

\bibitem{Qauli:2016PRD}
      A.I. Qauli, and A. Sulaksono,
         Phys. Rev D {\bf 93}, 025022 (2016).
% `` Quark matter at high density based on an extended confined isospin-density-dependent mass model."

\bibitem{Roberts:DSE-BSE}
     C. D. Roberts, and A. G. Williams,
         Prog. Part. Nucl. Phys. {\bf 33},  477 (1994);
% ``Dyson-Schwinger equations and their application to hadronic physics."
%
     C. D. Roberts, and S. Schmidt,
        Prog. Part. Nucl. Phys. {\bf 45}, S1 (2000);
% `` Dyson-Schwinger equations: Density, temperature and continuum strong QCD."
%
      P. Maris and C. D. Roberts,
         Int. J. Mod. Phys. E {\bf 12}, 297 (2003);
%  ``  Dyson-Schwinger equations: A tool for hadron physics."
%
      A. Bashir, L. Chang, I. C. Cloet, B. El-Bennich, Y. X. Liu, C. D. Roberts, and P. C. Tandy,
        Commun. Theor. Phys. {\bf 58}, 79 (2012).
% ``  Collective Perspective on Advances in Dyson-Schwinger Equation QCD  ".
%

\bibitem{McLerran:2007NPA}
      L. McLerran, and R. D. Pisarski,
         Nucl. Phys. A {\bf 796}, 83 (2007);
%  `` Phases of dense quarks at large Nc "
%
      K.L. Wang, S.X. Qin, Y. X. Liu, L. Chang, C. D. Roberts, and S. M. Schmidt,
        Phys. Rev. D {\bf 86}, 114001 (2012);
% ``Existence and stability of multiple solutions to the gap equation"
%
     S.X. Qin, and D.H. Rischke,
        Phys. Rev. D {\bf 88}, 056007 (2013).
% ``  Quark spectral function and deconfinement at nonzero temperature "

\bibitem{Qin:2011PRL}
      S.X. Qin, L. Chang, H. Chen, Y.X. Liu, and C. D. Roberts,
         Phys. Rev. Lett, {\bf 106}, 172301 (2011).
% ``Phase Diagram and Critical End Point for Strongly Interacting Quarks"

\bibitem{Gao:2016PRDa}
      F. Gao, J. Chen, Y. X. Liu, S. X. Qin, C. D. Roberts, and S. M. Schmidt,
        Phys. Rev. D {\bf 93}, 094019 (2016);
% ``Phase diagram and thermal properties of strong-interaction matter"
%
     F. Gao, and Y.X. Liu,
        {\bf 94}, 094030 (2016).
%  ``Interface effect in QCD phase transitions via Dyson-Schwinger equation approach"

\bibitem{Gao:2016PRDb}
      F. Gao, and Y. X. Liu,
        Phys. Rev. D {\bf 94}, 076009 (2016).
%  ``QCD phase transitions via a refined truncation of Dyson-Schwinger equations  "

\bibitem{QCDPT-DSE2}
      C. S. Fischer, J. Luecker, and J. A. Mueller,
          Phys. Lett. B {\bf 702}, 438 (2011);
% `` Chiral and deconfinement phase transitions of two-flavour QCD at finite temperature
%     and chemical potential "
%
      C. S. Fischer, and J. Luecker,
         Phys. Lett. B {\bf 718}, 1036 (2013);
%Propagators and phase structure of Nf= 2 and Nf= 2+ 1 QCD
%
     G. Eichmann, C. S. Fischer, and C. A. Welzbacher,
        Phys. Rev. D {\bf 93}, 034013 (2016).
% `` Baryon effects on the location of QCD's critical end point"

\bibitem{Chen:2011PRD}
     H. Chen, M. Baldo, G.F. Burgio, and H.-J. Schulze,
        Phys. Rev. D {\bf 84}, 105023 (2011).
% ``  Hybrid stars with the Dyson-Schwinger quark model. "

\bibitem{Chen:2012PRD}
      H. Chen, M. Baldo, G.F. Burgio, and H.-J. Schulze,
        Phys. Rev. D {\bf 86}, 045006 (2012).
%``  Hybrid protoneutron stars with the Dyson-Schwinger quark model "

\bibitem{Chen:2015PRD}
     H. Chen, J. -B. Wei, M. Baldo, G.F. Burgio, and H.-J. Schulze,
        Phys. Rev. D {\bf 91}, 105002 (2015).
% ``  Hybrid neutron stars with the Dyson-Schwinger quark model  and various quark-gluon vertices. "

\bibitem{Chen:2016EPJA}
     H. Chen, J. -B. Wei, and H. -J. Schulze,
        Eur. Phys. J. A {\bf 52}, 291 (2016).
% ``  Strange quark matter and quark stars with the Dyson-Schwinger quark model  "

\bibitem{Glendenning:1992PRD}
      N.K. Glendenning,
         Phys. Rev. D {\bf 46}, 1274 (1992).
% `` First-order phase transitions with more than one conserved charge: Consequences for neutron stars."

\bibitem{Maruyama:2007PRD}
      T. Maruyama, S. Chiba, H.-J. Schulze, and T. Tatsumi,
         Phys. Rev. D {\bf 76}, 123015 (2007).
% `` Hadron-quark mixed phase in hyperon stars."

\bibitem{Carroll:2009PRC}
      J.D. Carroll, D.B. Leinweber, A.G. Williams, and A.W. Thomas,
         Phys. Rev. C {\bf 79}, 045810 (2009).
% `` Phase transition from quark-meson coupling hyperonic matter to deconfined quark matter."

\bibitem{Weissenborn:2011ApJL}
      S. Weissenborn, I. Sagert, G. Pagliara, M. Hempel, and J. Schaffner-Bielich,
         Astrophys. J. {\bf 740}, L14 (2011).
% `` Quark Matter in Massive Compact Stars."

\bibitem{Fischer:2011ApJS}
      T. Fischer, et al.,
         Astrophys. J. Suppl. {\bf 194}, 39 (2011).
% ``  Core-collapse Supernova Explosions Triggered by a Quark-hadron phase transition
%   during the Early Post-bounce Phase."

\bibitem{Schulze:2011PRC}
      H.-J. Schulze, and T. Rijken,
         Phys. Rev. C {\bf 84}, 035801 (2011).
% `` Maximum mass of hyperon stars with the Nijmegen ESC08 model."

\bibitem{Wen:2013PA}
      X.J. Wen,
        Physica A {\bf 392}, 4388 (2013).
% ``  Equation of state in hybrid stars and the stability window of quark matter."

\bibitem{Xiao:2009PRL}
      Z.G. Xiao, B.A. Li, L.W. Chen, G.C. Yong, and M. Zhang,
        Phys. Rev. Lett. {\bf 102}, 062502 (2009).
%  ``  Circumstantial evidence for a soft nuclear symmetry energy at suprasaturation densities. "

\bibitem{Heiselberg:1993PRL}
	H. Heiselberg, C. J. Pethick, and E. F. Staubo,
		Phys. Rev. Lett. {\bf 70}, 1355 (1993).
%``Quark matter droplets in neutron stars''

\bibitem{Glendenning:1995PRC}
	N. K. Glendenning, and S. Pei,
	    Phys. Rev. C {\bf 52}, 2250 (1995).
%``Crystalline structure of the mixed confined-deconfined phase in neutron stars''

\bibitem{Christiansen:1997PRC}
	M. B. Christiansen, and N. K. Glendenning,
		Phys. Rev. C {\bf 56}, 2858 (1997).
%``Finite size effects and the mixed quark-hadron phase in neutron stars''


\bibitem{Endo:2006PTP}
	T. Endo, T. Maruyama, S. Chiba, and T. Tatsumi,
		Prog. Theor. Phys. {\bf 115}, 337 (2006).
%``Charge Screening Effect in the Hadron-Quark Mixed Phase''

\bibitem{Yasutake:2014PRC}
	N. Yasutake, R. Lastowiecki, S. Beni\ifmmode \acute{c}\else \'{c}\fi{}, D. Blaschke, T. Maruyama, and T. Tatsumi,
		Phys. Rev. C {\bf 89}, 065803 (2014).
%``Finite-size effects at the hadron-quark transition and heavy hybrid stars''
	
\bibitem{Wu:2017PRC}
	X. H. Wu, and H. Shen,
		Phys. Rev. C {\bf 96}, 025802 (2017).
%``Finite-size effects on the hadron-quark phase transition in neutron stars''

\bibitem{Masuda:2013ApJ}
      K. Masuda, T. Hatsuda, and T. Takatsuka,
         Astrophys. J. {\bf 764}, 12 (2013).
% `` Hadron-quark crossover and massive hybrid stars with strangeness. "

\bibitem{Masuda:2013PTEP}
      K. Masuda, T. Hatsuda, and T. Takatsuka,
         Prog. Theor. Exp. Phys. {\bf 7}, 073D01 (2013).
% `` Hadron-quark crossover and massive hybrid stars.  "

\bibitem{Kojo:2015PRD}
      T. Kojo, P.D. Powell, Y. Song, and G. Baym,
         Phys. Rev. D {\bf 91}, 045003 (2015).
% ``  Phenomenological QCD equation of state for massive neutron stars."

\bibitem{Kojo:2016EPJA}
      T. Kojo,
         Eur. Phys. J. A {\bf 52}, 51 (2016).
% `` Phenomenological neutron star equations of state 3-window modeling of QCD matter. "

\bibitem{Baym:2017RPP}
      G. Baym, T. Hatsuda, T. Kojo, P.D. Powell, Y. Song, and T. Takatsuka,
 %        Rep. Prog. Phys. {\bf ??}, ?????? (2017).
       arXiv:1707.04966 (2017).
% ``  From hadrons to quarks in neutron stars."

\bibitem{Walecka:1974AP}
      J. D. Walecka,
        Ann. Phys. {\bf 83}, 491 (1074).
% `` A theory of highly condensed matter. "

\bibitem{Boguta:1977}
      J. Boguta, and A. R. Rodmir,
        Nucl. Phys. A {\bf 292}, 413 (1977);
% `` Relativistic calculation of nuclear matter and the nuclear surface. "
      J. Boguta,
        Phys. Lett.  {\bf 106B}, 255 (1981);
% `` Remarks on the Beta stability in neutron stars. "
      J. Boguta, and H. Stocker,
        Phys. Lett.  {\bf 120B}, 289 (1983);
% `` Systematics of nuclear matter properties in a non-linear relativistic field theory.  "
      W. Pannert, P. Ring, and J. Boguta,
        Phys. Rev. Lett. {\bf 59}, 2420 (1987);
% `` Relativistic mean-field theory and nuclear deformation.  "
      J. Boguta,
        Nucl. Phys. A {\bf 501}, 637 (1989).
% `` Chiral nuclear interaction. "

\bibitem{Serot:1986ANP}
      B. D. Serot, and J. D. Walecka,
%  ``The Relativistic Nuclear Many Body Problem",
         Adv. Nucl. Phys.  {\bf 16},  1 (1986).


\bibitem{Dutra:2014PRC}
      M. Dutra, O. Louren\ifmmode \mbox{\c{c}}\else\c{c}\fi{}o, S.S. Avancini, B.V.Carlson,
	   A. Delfino, D.P. Menezes, C. Provid\^encia, S. Typel, and J.R. Stone,
         Phys. Rev. C {\bf 90}, 055203 (2014).
% `` Relativistic mean-field hadronic models under nuclear matter constraints . "

\bibitem{Dutra:2016PRC}
      M. Dutra, O. Louren\ifmmode \mbox{\c{c}}\else\c{c}\fi{}o, and D.P. Menezes,
         Phys. Rev. C {\bf 93}, 025806 (2016).
% ``  Stellar properties and nuclear matter constraints. "

\bibitem{Typel:1999NPA}
      S. Typel, and H.H. Wolter,
          Nucl. Phys. A {\bf 656}, 331 (1999).
% `` Relativistic mean field calculations with density-dependent meson-nucleon coupling ."

\bibitem{Kolomeitsev:2017NPA}
	E. E. Kolomeitsev, K. A. Maslov, D. N. Voskresensky,
		Nucl. Phys. A {\bf 961}, 106 (2017).
%``Delta isobars in relativistic mean-field models with sigma-scaled hadron masses and couplings"

\bibitem{Maris:1997PRC}
      P. Maris, and C.D. Roberts,
         Phys. Rev. C {\bf 56}, 3369 (1997).
% ``  Pi and K-meson Bethe-Salpeter amplitudes  "

\bibitem{Alkofer:2002PRD}
      R. Alkofer, W. Detmold, C.S. Fischer, and P. Maris,
         Phys. Rev. D {\bf 70}, 014014 (2004).
% `` Analytic properties of the Landau gauge gluon and quark propagators ."
%
\bibitem{Chang:2005NPA}
      L. Chang, Y. X. Liu, and H. Guo,
         Nucl. Phys. A {\bf 750}, 324 (2005).
% `` Density dependence of nucleon radius and mass in the global color symmetry model of QCD with
%  a sophisticated effective gluon propagator  ."

\bibitem{Chen:2008PRD}
      H. Chen, W. Yuan, L. Chang, Y.X. Liu, T. Kl\"ahn, and C.D. Roberts,
         Phys. Rev. D {\bf 78}, 116015 (2008).
% ``  Chemical potential and the gap equation."

\bibitem{Klahn:2010PRC}
      T. Kl\"ahn, C.D. Roberts, L. Chang, H. Chen, and Y.X. Liu,
         Phys. Rev. C {\bf 82}, 035801 (2010).
% ``  Cold quarks in medium: An equation of state."

\bibitem{Haymaker:1991RNCSIF}
      R.W. Haymaker,
         Riv. Nuovo Cimento Soc. Ital. Fis. {\bf 14}, 1 (1991).
%  ``  Variational-methods for Composite-Operators."

\bibitem{Zhu:2016PRC}
	Z.Y. Zhu, A. Li, J. N. Hu, H. Sagawa,
		Phys. Rev. C {\bf 94}, 045803 (2016)
		%`` Delta(1232) effects in density-dependent relativistic Hatree-Fock theory and neutron stars.

\bibitem{Margalit:2017}
	B. Margalit, and B. D. Metzger,
	arXiv:1710.05938 (2017)
%``Constraining the maximum mass of neutron stars from multi-messenger observations of GW170817''

\bibitem{Hebeler:2010PRL}
      K. Hebeler, J.M. Lattimer, C.J. Pethick, and A. Schwenk,
          Phys. Rev. Lett. {\bf 105}, 161102 (2010).
% ``  Constraints on Neutron Star Radii Based on Chiral Effective Field Theory Interactions."

\bibitem{Guillot:2014ApJL}
      S. Guillot, and R.E. Rutledge,
         Astrophys. J. {\bf 796}, L3 (2014).
% ``  Rejecting proposed dense matter equation of state with quiescent low-mass X-ray binaries ."

\bibitem{Bauswein:2017}
	A. Bauswein, O. Just, H.-T. Janka, and N. Stergioulas,
	arXiv:1710.06843, (2017)
%``Neutron-star radius constraints from GW170817 and future detections''

\bibitem{Liu:20013NPA}
      Y.X. Liu, D.F. Gao, and H. Guo,
         Nucl. Phys. A {\bf 695}, 353 (2001);
% `` Density dependence of nucleon bag constant, radius and mass in an effective field theory model
%   of QCD ."
      Y.X. Liu, D.F. Gao, J. H. Zhou, and H. Guo,
         Nucl. Phys. A {\bf 725}, 127 (2003).
% ``  Reevaluation of the density dependence of nucleon radius and mass
%   in the global color symmetry model of QCD."
%

\bibitem{Liu:2010PRC}
      Y. Mo, S.X. Qin, and Y.X. Liu,
         Phys. Rev. C {\bf 82}, 025206 (2010).
% ``  Temperature dependence of the effective bag constant and the radius of a nucleon
%   in the global color symmetry model of QCD."
%

\bibitem{Wang:2013PRD}
      K.L. Wang, Y. X. Liu, L. Chang, C. D. Roberts, and S. M. Schmidt,
         Phys. Rev. D {\bf 87}, 074038 (2013).
% ``  Baryon and meson screening masses."

\end{thebibliography}
%

\end{document}